\documentclass[fleqn,10pt]{wlscirep}
\usepackage[utf8]{inputenc}
\usepackage[T1]{fontenc}

\usepackage{graphicx}
\usepackage{amssymb}
\usepackage{multirow}
\usepackage[caption=false]{subfig} 
\usepackage{pifont}
\usepackage{placeins}
\usepackage{amsmath} 
\usepackage{url}  
\usepackage{cite}
\usepackage{rotating}
\DeclareMathOperator{\Tr}{Tr} 

\title{Detecting coalitions by optimally partitioning signed networks of political collaboration
} 

\author[1,2*]{Samin Aref}
\author[3]{Zachary Neal}
\affil[1]{Laboratory of Digital and Computational Demography, Max Planck Institute for Demographic Research, Germany}
\affil[2]{School of Computer Science, University of Auckland, New Zealand}
\affil[3]{Department of Psychology, Michigan State University, USA}
\affil[*]{sare618@aucklanduni.ac.nz}

\begin{abstract}
    We propose new mathematical programming models for optimal partitioning of a signed graph into cohesive groups. To demonstrate the approach's utility, we apply it to identify coalitions in US Congress since 1979 and examine the impact of polarized coalitions on the effectiveness of passing bills. Our models produce a globally optimal solution to the NP-hard problem of minimizing the total number of intra-group negative and inter-group positive edges. We tackle the intensive computations of dense signed networks by providing upper and lower bounds, then solving an optimization model which closes the gap between the two bounds and returns the optimal partitioning of vertices. Our substantive findings suggest that the dominance of an ideologically homogeneous coalition (i.e.\ partisan polarization) can be a protective factor that enhances legislative effectiveness.
  The reference to this article should be made as follows: {\scshape Aref, S., and Neal, Z.}
  	\newblock Detecting coalitions by optimally partitioning signed networks of political collaboration.
  	\newblock {\em Scientific Reports}, (2020),
  	\newblock \url{www.doi.org/10.1038/s41598-020-58471-z}.\\
  	
  	\textbf{Old title:} Legislative effectiveness hangs in the balance: Studying balance and polarization through partitioning signed networks
  
\end{abstract}

\begin{document}

\flushbottom
\maketitle


\thispagestyle{empty}
\section*{Introduction}

We propose a general method for identifying cohesive groups in signed networks (networks with positive and negative edges), and apply it to political networks, which have become a common focus in complex network analysis\cite{robero2018, colliri2019, faustino2019}. Specifically, we examine signed networks of political collaboration and opposition to identify the members of polarized coalitions in the US Congress, then use these coalitions to examine the impact of polarization on effectiveness in passing bills.

In legislative bodies where most pairs of legislators co-sponsor bills, a network of who co-sponsors with whom becomes a highly dense network which would not be suitable for studying political alliances, coalitions, and polarization. Instead, we use signed networks \cite{Neal2019figshare} created based on significantly \textit{many} and significantly \textit{few} co-sponsorships as two types of edges with opposite nature where a stochastic degree sequence model (SDSM) \cite{neal_backbone_2014} is used as the null model, to define thresholds of “many” and “few.”
Previous research \cite{neal2018} on the same data has shown an increase in polarization in the US Congress when measured by the triangle index, which provides a locally-aggregated index of polarization based on structural balance \cite{terzi_spectral_2011}. However, the triangle index only measures the level of balance and polarization, but does not identify the members of the political coalitions that are polarized. For this we turn to the frustration index \cite{zaslavsky_balanced_1987,facchetti_computing_2011,aref2015measuring} (also known as the line index of balance \cite{harary_measurement_1959}), which optimally partitions a signed graph into two opposing but internally cohesive ``coalitions'' \cite{harary_simple_1980}. Substantively, these coalitions \cite{riker1962theory} represent groups of legislators who sponsor significantly many bills with each other (i.e.\ are political allies), but who sponsor significantly few bills with those in the other coalition (i.e.\ are political enemies). In our analyses of legislative effectiveness, we focus on the level of partisanship within the largest, and therefore controlling, coalition.

Computing the frustration index is an NP-hard problem \cite{huffner_separator-based_2010}, and so is the equivalent partitioning problem that deals with minimizing the total number of intra-group negative and inter-group positive edges. The optimality of a numerical solution to an instance of an optimization problem depends on the function under optimization. Most studies on this topic use heuristic methods for partitioning signed networks under similar objectives \cite{gong_network_2017,traag_partitioning_2018,hua_fast_2018,brusco_partitioning_2019}. These methods are not guaranteed to provide the optimal solution or even its approximation within a constant factor \cite{huffner_separator-based_2010,aref2016exact}, but can potentially be implemented on larger networks.

Computing the exact value of the frustration index, in principle, involves searching among all possible ways to partition a given signed network into $k \leq 2$ groups in order to find the partitioning which minimizes the total number of intra-group negative and inter-group positive edges. We propose a new method for tackling the intensive computations by providing upper and lower bounds for this number, then solving an optimization model which closes the gap between the two bounds and returns the exact value of frustration index alongside the optimal partitioning of vertices.

\section*{Signed graph and balance theory preliminaries} \label{s:preliminaries}

In this section, we recall some basic definitions of signed graphs and balance theory.

\subsection*{Signed graphs}
\label{ss:signedgraph}
We consider an undirected signed graph $G = (V,E,\sigma)$ where $V$ and $E$ are the sets of vertices and edges respectively, and $\sigma$ is the sign function $\sigma: E\rightarrow\{-1,+1\}$. Graph $G$ contains $|V|=n$ nodes. The set $E$ of edges contains $m^-$ negative edges and $m^+$ positive edges adding up to a total of $|E|=m=m^+ + m^-$ edges. 
The \emph{signed adjacency matrix} and the \emph{unsigned adjacency matrix} are denoted by $\textbf{A}$ and ${|\textbf{A}|}$ respectively. Their entries are defined in Eq.\ \eqref{eq1} and Eq.\ \eqref{eq1.1}. 
\begin{align}\label{eq1}
a{_u}{_v} =
\left\{
\begin{array}{ll}
\sigma_{(u,v)} & \quad \mbox{if } {(u,v)}\in E \\
0 & \quad \mbox{if } {(u,v)}\notin E
\end{array}
\right.
\end{align}
\begin{align}\label{eq1.1}
|a{_u}{_v}| =
\left\{
\begin{array}{ll}
1 & \quad \mbox{if } {(u,v)}\in E \\
0 & \quad \mbox{if } {(u,v)}\notin E
\end{array}
\right.
\end{align}

\subsection*{Balance and cycles}
\label{subsec:balance}
A \emph{cycle} of length $k$ in $G$ is a sequence of nodes $v_0,v_1,...,v_{k-1},v_k$ such that for each $i=1,2,...,k$ there is an edge from $v_{i-1}$ to $v_i$ and the nodes in the sequence except for $v_0=v_k$ are distinct. The \emph{sign} of a cycle is the product of the signs of its edges. A cycle with negative (positive) sign is unbalanced (balanced). A balanced network (graph) is one with no negative cycles.

Balance theory is conceptualized by Heider in the context of social psychology \cite{heider_social_1944}. It was then formulated as a set of graph-theoretic conditions by Cartwright and Harary \cite{cartwright_structural_1956} which define a signed graph to be balanced if all its cycles are positive. Cartwright and Harary also introduce measuring the level of balance using, among other indices, the fraction of positive cycles \cite[page 288]{cartwright_structural_1956}. Three years later, Harary suggested using frustration index \cite{harary_measurement_1959} (under a different name); a measure which satisfies key axiomatic properties \cite{aref2015measuring}, but has been underused for decades due to the complexity involving its computation \cite{aref2017computing,aref2016exact,aref2017balance}. 

\section*{Evaluating balance and frustration}
\label{sec:evaluate}

In this section, we explain our computational approach to analyzing signed networks by providing brief definitions and discussions on measuring balance, frustration and partitioning, and graph optimization models.

\subsection*{Measuring partial balance}
\label{subsec:methodology}
Signed networks representing real data are often unbalanced, which motivates measuring the intermediate level of partial balance \cite{aref2015measuring}. The first measure we use is \emph{triangle index} denoted by $T(G)$ which equals the fraction of positive cycles of length 3 \cite{cartwright_structural_1956,harary1977graphing}. We use Eq.\ \eqref{eq1.8} suggested in \cite{terzi_spectral_2011} for computing triangle index, $T(G)$, in which $\Tr(\textbf{A})$ denotes the trace (sum of diagonal entries) of ${\textbf{A}}$. 
\begin{align}\label{eq1.8}
T(G) = \frac{\Tr({\textbf{A}}^3)+\Tr({|\textbf{A}|}^3)}{2 \Tr({|\textbf{A}|}^3)} 
\end{align}

The other measure we use is the \textit{normalized frustration index} \cite{aref2015measuring} denoted by $F(G)$ which is based on normalizing the minimum number of edges whose removal results in a balanced graph \cite{abelson_symbolic_1958,harary_measurement_1959,zaslavsky_balanced_1987}. 

\begin{figure}[ht]
	\centering
	\includegraphics[width=0.9\textwidth]{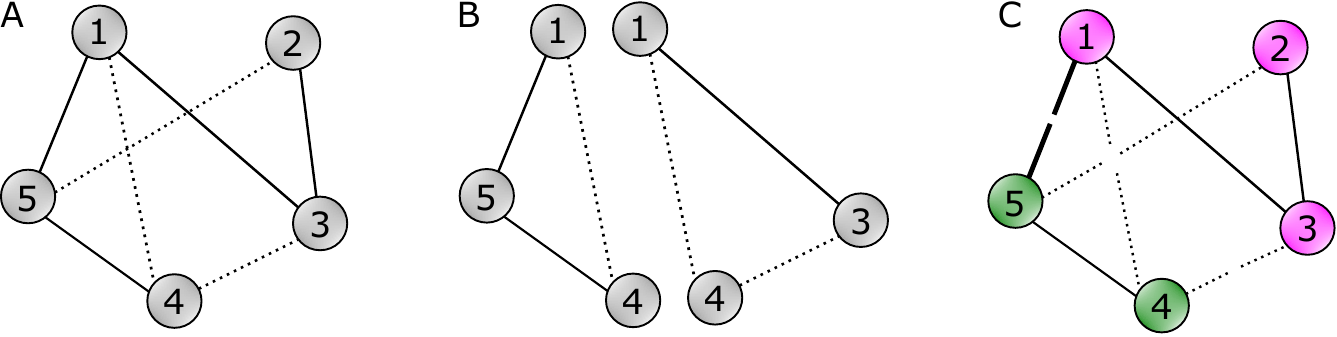}
	\caption{(A) An example signed network. (B) Evaluating balance using triangles. (C) Evaluating balance using frustration.}
	\label{fig:toy}
\end{figure}

Fig. \ref{fig:toy}(A) shows an example signed graph in which the three dotted lines represent negative edges and the four solid lines represent positive edges. The level of balance in this signed graph can be evaluated using triangles (B) or frustration (C). The former approach, (B), involves identifying triangle 1-4-5 as unbalanced and triangle 1-3-4 as balanced leading to the numeric index $T(G)=1/2$. The latter approach, (C), involves finding a partitioning of vertices $\{\{1,2,3\},\{4,5\}\}$ (shown by green and purple colors in Fig. \ref{fig:toy}) which minimizes the total number of intra-group negative and inter-group positive edges to $1$ (only edge $(1,5)$ according to this partitioning). Note that removing edge $(1,5)$ leads to a balanced signed graph.

\subsection*{Frustration and partitioning}
\label{subsec:frustration}
Given signed graph $G=(V,E,\sigma)$, we can partition $V$ into two subsets: $X$ and $V\setminus X$. We let binary variable $x_i$ denote the subset which node $i$ belongs to under partitioning $\{X,V\setminus X\}$, where $x_i=1$ if $i \in X$ and $x_i=0$ otherwise. 

A positive edge $(i,j) \in E^+$ is said to be frustrated if its endpoints $i$ and $j$ belong to different subsets ($x_i \ne x_j$). A negative edge $(i,j) \in E^-$ is said to be frustrated if its endpoints $i$ and $j$ belong to the same subset ($x_i = x_j$). 
We define the {\em frustration count} $f_G(X)$ as the number of frustrated edges of $G$ under partitioning $\{X,V\setminus X\}$: $$f_G(X) = \sum_{(i,j) \in E} f_{ij}(X)$$
where $f_{ij}(X)$ is the frustration state of edge $(i,j)$, given by
\begin{equation} \label{eq2}
f_{ij}(X)=
\begin{cases}
0, & \text{if}\ x_i = x_j \text{ and } (i,j) \in E^+ \\
1, & \text{if}\ x_i = x_j \text{ and } (i,j) \in E^- \\
0, & \text{if}\ x_i \ne x_j \text{ and } (i,j) \in E^- \\
1, & \text{if}\ x_i \ne x_j \text{ and } (i,j) \in E^+. \\
\end{cases}
\end{equation}

The frustration index of a graph $G$ can be computed exactly by finding partitioning $X^*, V\setminus X^* \subseteq V$ of $G$ that minimizes the frustration count $f_G(X)$, i.e.\, solving  Eq.\ \eqref{eq3} \cite{aref2017computing,aref2016exact}. 
\begin{equation} \label{eq3}
L(G) = \min_{X \subseteq V}f_G(X)\
\end{equation}

The normalized frustration index, $F(G)$, is computed based on $L(G)$ and according to Eq.\ \eqref{eq1.95} which allows measuring the level of partial balance based on numerical values within the unit interval ($m$ denotes the number of edges).
\begin{align}\label{eq1.95}
F(G)=1-2L(G)/m
\end{align}

One may notice some similarities between the problem of finding communities in unsigned networks \cite{KL1970,
GN2002,CNM2004,Raghavan2007,Cordasco2010,Pares2017} and that of partitioning signed networks to minimize the frustration count. One key difference is that in the latter problem for every pair of vertices there are three cases (as opposed to two): a positive edge, a negative edge, or no edge between the two vertices. Due to the differences between objectives of these two problems (minimizing frustration count as opposed to maximizing modularity or other quantities), the partitioning obtained from running community detection algorithms on positive edges of a signed graph will not generally minimize the frustration count.

Recent studies on frustration index and signed networks suggest \cite{aref2017computing,aref2016exact} and implement \cite{aref2017balance} efficient graph optimization models to compute the frustration index of relatively large (up to $10^5$ edges) sparse networks. However, the signed networks we analyze have substantially higher densities compared to the instances in \cite{aref2017computing,aref2016exact,aref2017balance}. This requires developing new computational models for tackling the intensive computations involved in obtaining the frustration index of dense graphs. 

\subsection*{Bounding the frustration index}
\label{subsec:bound}

In this subsection, we discuss obtaining lower and upper bounds for the frustration index. Using these bounds is a way of substantially reducing the running time, but theoretically they are not required.

The linear programming relaxation (LP relaxation) of the binary optimization models in \cite{aref2017computing,aref2016exact} can be used to compute a lower bound for the frustration index. The linear programming model in Eq.\ \eqref{eq3.6} is developed for this purpose. 
\begin{equation}
\label{eq3.6}
\begin{split}
\min_{x_i, x_{ij}} Y = &\sum\limits_{(i,j) \in E^+} x_{i} + x_{j} - 2x_{ij} + \sum\limits_{(i,j) \in E^-} 1 - (x_{i} + x_{j} - 2x_{ij})\\
\text{s.t.} \quad
x_{ij} &\leq (x_{i}+x_{j})/2 \quad \forall (i,j) \in E^+ \\
x_{ij} &\geq x_{i}+x_{j}-1 \quad \forall (i,j) \in E^- \\
x_{i}+x_{jk} &\geq x_{ij} + x_{ik} 										  \quad \forall (i,j,k) \in T\\
x_{j}+x_{ik} &\geq x_{ij} + x_{jk}  											\quad \forall (i,j,k) \in T\\
x_{k}+x_{ij} &\geq x_{ik} + x_{jk}  											\quad \forall (i,j,k) \in T\\
1 + x_{ij} + x_{ik} + x_{jk} &\geq x_{i} + x_{j} + x_{k} \quad \forall (i,j,k) \in T \\
x_{i} &\in [0,1] \quad  \forall i \in V \\
x_{ij} &\in [0,1] \quad \forall (i,j) \in E 
\end{split}
\end{equation}

In Eq.\ \eqref{eq3.6}, $T=\{(i,j,k)\in V^3 \mid (i,j),(i,k),(j,k) \in E \}$ is the set which contains ordered 3-tuples of nodes whose edges form a triangle in $G$. The continuous linear programming model in Eq.\ \eqref{eq3.6} is developed by combining the LP relaxation of the 0/1 linear model in \cite{aref2017computing}[Subsection 4.3] and the triangle constraints in \cite{aref2017computing}[Subsection 4.4]. It follows from the LP relaxation that the optimal solution $Y^*$ to the model in Eq.\ \eqref{eq3.6} is a lower bound for the frustration index $Y^* \leq L(G)$.

Any given partitioning $\{X,V\setminus X\}$ for signed graph $G$ is associated with a frustration count $f_G(X)$ which is by definition (as in Eq.\ \eqref{eq3}) an upper bound for the frustration index $$f_G(X^*) = L(G) \leq f_G(X) \quad \forall X \subseteq V.$$

We use a specific partitioning $\{X',V\setminus X'\}$ as a starting point to ``warm-start'' the algorithm for computing the frustration index. Partitioning $\{X',V\setminus X'\}$ groups nodes into two subsets based on the party affiliation of legislators. To be more precise, for node $i$ which represents a legislator, decision variable $x_{i}$ is given initial value $0$ if the reciprocal legislator is a Democrat and $x_{i}$ is given initial value $1$ otherwise.

\subsection*{Computing the frustration index}
\label{subsec:compute}

After bounding the frustration index, we use the binary linear programming model in Eq.\ \eqref{eq4} which minimizes the number of frustrated edges.
\begin{equation}
\label{eq4}
\begin{split}
\min_{x_i, f_{ij}} Z &= \sum\limits_{(i,j) \in E}  f_{ij}  \\
\text{s.t.} \quad
f_{ij} &\geq x_{i}-x_{j} \quad \forall (i,j) \in E^+ \\
f_{ij} &\geq x_{j}-x_{i} \quad \forall (i,j) \in E^+ \\
f_{ij}  &\ge  x_{i} + x_{j} -1 \quad \forall (i,j) \in E^- \\
f_{ij}  &\ge  1-x_{i} - x_{j}  \quad \forall (i,j) \in E^- \\
\sum\limits_{(i,j) \in E}  f_{ij} &\geq Y^* \\
x_{i} &\in \{0,1\} \quad  \forall i \in V \\
f_{ij} &\in \{0,1\} \quad \forall (i,j) \in E 
\end{split}
\end{equation}

The binary variables of the model are $f_{ij} \: \forall (i,j) \in E$ which denotes frustration of edge $(i,j)$ and $x_i \: \forall i \in V$ which denotes the subset of node $i$. To warm-start the algorithm which solves Eq.\ \eqref{eq4}, we initialize $x_i$ variables based on partitioning $\{X',V\setminus X'\}$. The model in Eq.\ \eqref{eq4} is developed by combining the XOR model in \cite{aref2016exact}[Subsection 3.2] with an additional constraint to incorporate the lower bound $Y^*$ obtained from Eq.\ \eqref{eq3.6}. We implement the speed-up techniques discussed in \cite{aref2016exact} 
and solve the binary linear programming model in Eq.\ \eqref{eq4} using \textit{Gurobi} solver (version 8.0) \cite{gurobi} on a virtual machine with 32 Intel Xeon CPU E5-2698 v3 @ 2.30 GHz processors and 32 GB of RAM running 64-bit Microsoft Windows Server 2012 R2 Standard. 

\section*{Results}
\label{sec:result}
In this section, we provide the results of analyzing balance and frustration in signed networks of US Congress legislators.

\subsection*{Partial balance, frustration, and optimal partitioning}
\label{subsec:result1}
\begin{figure}[ht]
	\centering
		\includegraphics[width=1\textwidth]{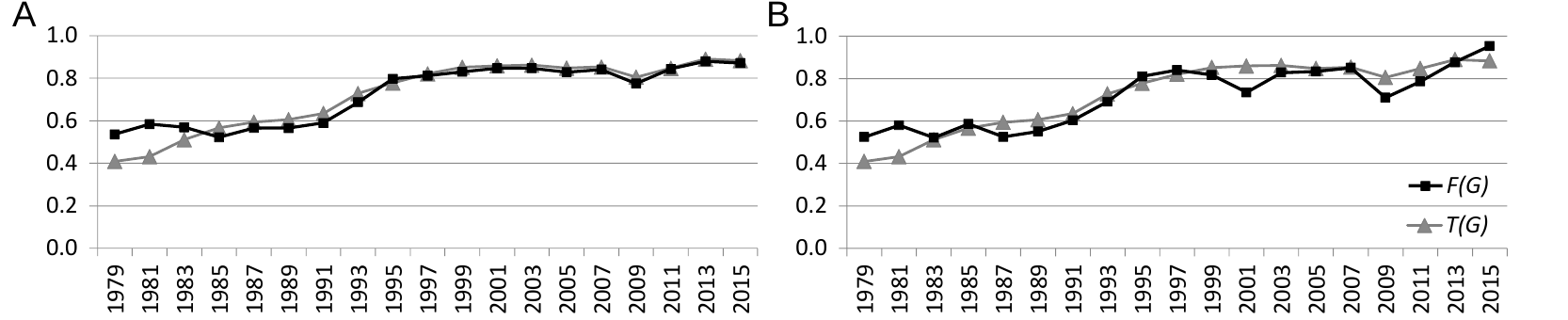}
	\caption{Two measures of partial balance indicating an overall increase in political polarization in (A) US House of Representatives and (B) US Senate over the time period 1979-2016}
	\label{fig:balance}
\end{figure}

We evaluate the level of partial balance using two different methods. Fig. \ref{fig:balance} illustrates partial balance in the signed networks of the US Congress over time measured by the triangle index and normalized frustration index. 
Values of the two measures, the triangle index $T(G)$ and the normalized frustration index $F(G)$, are highly correlated (correlation coefficients are $0.95$ and $0.91$ respectively for House and Senate networks) and both show relatively high levels of partial balance which have increased in the time period 1979-2016. The results in Fig. \ref{fig:balance} indicate an increase in the polarization of both chambers of US Congress, which is in accordance with the literature \cite{layman2006,zhang2008,waugh2011,moody2013,neal2018}. Although the triangle index $T(G)$ and the normalized frustration index $F(G)$ capture very similar information concerning the level of partial balance, only the computation of $F(G)$ also provides the partitioning that minimizes the sum of intra-group negative and inter-group positive edges.

Solving the continuous optimization model in Eq.\ \eqref{eq3.6} and the discrete (binary) optimization model in Eq.\ \eqref{eq4} requires intensive computations for large instances such as signed networks of the House. Given the size and density of these instances, the models in Eq.\ \eqref{eq3.6} and Eq.\ \eqref{eq4} have thousands of variables and possibly millions of constraints requiring a high performance computer taking advantage of parallel computing capabilities \cite{aref2017balance}.

For example, the signed graph instance of the 113th House session has $m=75,771$ edges and $|T|=7,102,625$ triangles which result in a total of $m+4|T|=28,486,271$ constraints for the model in Eq.\ \eqref{eq3.6}. Gurobi solver takes $5300$ seconds (around 1.5 hours) to solve the model in Eq.\ \eqref{eq3.6} to global optimality and return $Y^*$, the lower bound for the frustration index. For the same instance, the discrete optimization model in Eq.\ \eqref{eq4} has $n+m= 76,218$ binary variables and $2m+1=151,543$ constraints. This large instance takes $43,523$ seconds (around 12 hours) for Gurobi to reach global optimality and return the frustration index and the partitioning of the nodes. In total for the 113th session of the House, it takes $48,823$ seconds (around 13.5 hours) to compute the exact value of the frustration index which is the longest solve time among all instances. The average computation time for House instances is $17,763$ seconds (around 5 hours) and the standard deviation is $15,411$ seconds (around 4.5 hours). For Senate instances, the average computation time is 4 seconds and the standard deviation is 6 seconds.


Using the optimal values of the $x_i$ variables obtained by solving the discrete optimization model in Eq.\ \eqref{eq4}, we partition nodes of each network into two groups (subsets $X^*$, $V \setminus X^*$). For each signed network, either $X^*$ or $V \setminus X^*$ has the larger set cardinality and therefore represents the largest coalition for the corresponding session. 

\begin{figure}[ht]
	\centering
		\includegraphics[width=1\textwidth]{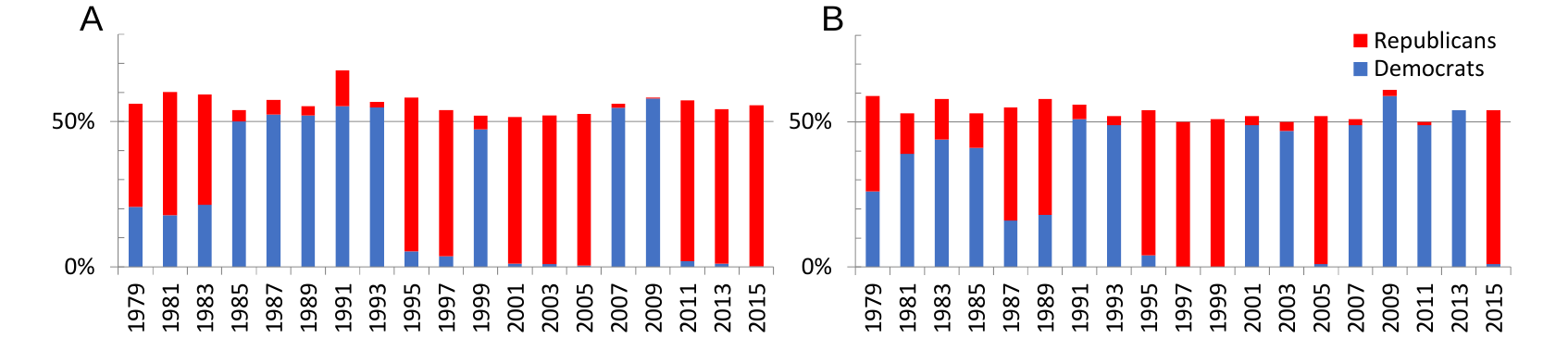}
	\caption{The number of legislators from the two main parties who belong to the controlling coalition indicating an increase in the partisan homogeneity of the controlling coalition in (A) US House of Representatives and (B) US Senate over the time period 1979-2016}
	\label{fig:composition}
\end{figure}

We evaluate the composition of the largest and therefore controlling coalitions in each session and chamber based on the party affiliation of its legislators. Fig. \ref{fig:composition} illustrates the number of legislators from the two main political parties in the controlling coalitions of the US Congress. As it can be seen in Fig. \ref{fig:composition}, the controlling coalitions have become more homogeneous (i.e.\ partisan) over the time period 1979-2016.

\subsection*{Legislative effectiveness and polarization in the US congress}
\begin{figure}[ht]
\centering
		\includegraphics[width=1\textwidth]{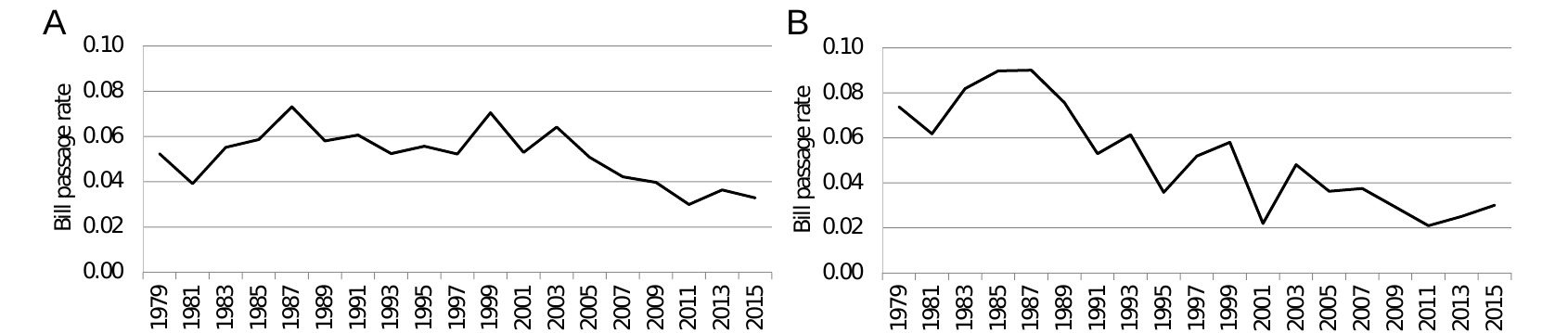}
	\caption{The fraction of bills that eventually become law (bill passage rate) indicating legislative effectiveness in (A) US House of Representatives and (B) US Senate over the time period 1979-2016}
	\label{fig:bill}
\end{figure}

Within the field of comparative US politics, two topics attract particular attention at the federal level: legislative effectiveness and political polarization. Legislative effectiveness refers to the ability of individual legislators \cite{olson_measures_1972,frantzich_who_1979}, or of an entire legislative body \cite{volden2014}, to advance their agenda, typically by facilitating the passage of legislation. Political polarization (when applied to elected officials or ``elites'') refers to the formation of non-overlapping ideologically homogeneous groups \cite{layman2006,neal2018}. When these groups mirror political party affiliations, 
it is also called partisan polarization. For several decades, legislative effectiveness in the US has declined (as illustrated in Fig. \ref{fig:bill}), while partisan polarization has increased \cite{neal2018}. These trends have led many to hypothesize that they are related, and specifically that ``unified party control has [not] been legislatively more productive than divided party control'' \cite[xii]{mayhew2005}.
Based on the legislative process used by the US Congress, it might be expected that a chamber's bills are more likely to become law when the controlling party holds a larger majority, because its members can form a voting bloc. However, the analysis in the next section suggests that that changes in bill passage rates are better explained by the partisanship of a chamber's largest coalition.

\subsection*{Mediation in bill passage}
\label{subsec:result3}

\begin{figure}[ht]
\centering
	\includegraphics[width=0.8\textwidth]{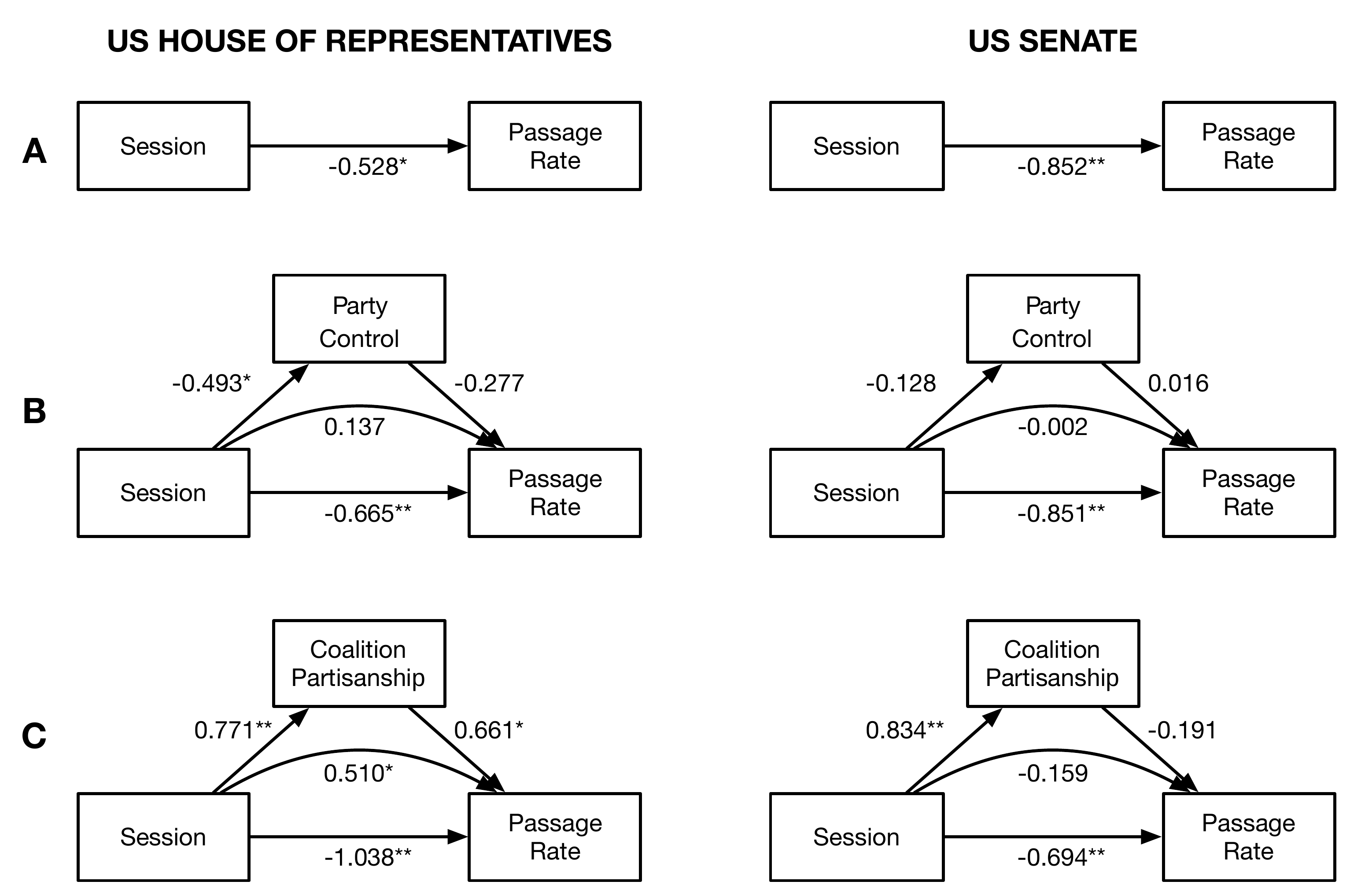}
	\caption{Predicting the rate of bill passage in the US House of Representatives and Senate; Standardized coefficients are reported; $** p < .01, * p < .05$. (A) Over time, the passage rate has declined. (B) The decline is not mediated by changes in the size of a party's majority in a chamber. (C) In the House, but not the Senate, it is mediated by the partisan homogeneity in the controlling coalition.}
	\label{fig:mediation}
\end{figure}

Using a bivariate linear regression, we find that the percentage of bills introduced in a chamber that become law (passage rate) significantly declines over time. The passage rate has declined in the House by an average of $0.11$ percentage points each session ($\beta = -0.528, p < 0.05$), and in the Senate by an average of $0.35$ percentage points each session ($\beta = -0.852, p < 0.01$; see Fig. \ref{fig:mediation}(A)). 
We report standardized coefficients ($\beta$) to facilitate cross-model comparisons. The percentage point changes are unstandardized bivariate regression coefficients, reported here for context. These variations in passage rate are not simply a function of the total number of bills introduced (House: $r = -0.29, p = 0.22$; Senate: $r = -0.08, p = 0.74$), which exhibit no association with time (House: $r = -0.34, p = 0.15$; Senate: $r = 0.19, p = 0.43$; see tables S2 and S4).

To investigate possible explanations for variations in passage rate, we estimate two separate structural equation models for each chamber. A commonsense model tests the expectation that when the majority party holds a larger numerical majority, they should have greater success passing bills \cite{moore_legislative_1969}. The key variable in this model, \emph{party control}, is defined as the absolute difference between the number of Republicans and Democrats. Computing party control does not require any information about the legislators' network. We find no support for this model; party control does not mediate the relationship between time and passage rate. (see Fig. \ref{fig:mediation}(B)).

A more nuanced model tests the expectation that when the controlling coalition is more partisan and thus more ideologically unified, it will have greater success passing bills. The key variable in this model, \emph{coalition partisanship}, is defined as the fraction of non-independent members in the largest coalition that affiliate with that coalition's dominant political party (see Fig. \ref{fig:composition}).
We compute coalition partisanship by applying the partition method described above to a signed network of political collaboration and opposition. We find support for this model in the US House, but not the US Senate (see Fig. \ref{fig:mediation}(C)). Specifically, we find that in the House, the partisan homogeneity of the controlling coalition has increased over time ($\beta = 0.771, p < .01$), which is consistent with past findings concerning polarization \cite{neal2018}, and that coalition homogeneity increases passage rates ($\beta = 0.661, p < 0.05$), which is consistent with our expectation about the impact of ideological unity. Together, these effects imply a significant and positive indirect effect of time on the passage rate ($\beta = 0.510, p < .05$), mediated by coalition partisanship. Thus, the observed decline in bill passage rates in the US House would have been worse (direct effect: $\beta = -1.038, p < .01$), but was mitigated by increasingly ideologically homogeneous coalitions, which are a protective factor against declines in legislative effectiveness.

\section*{Summary and conclusions}
\label{sec:conlcusion}
In this study we proposed a general method for identifying internally cohesive opposing coalitions in signed networks of legislators based on structural balance theory, then applied this method to identify opposing coalitions in the US Congress, showing that these coalitions' partisanship can explain changes in legislative effectiveness better than political parties. Based on this analysis, we offer a series of substantive and methodological conclusions.

Consistent with prior studies \cite{layman2006,zhang2008,waugh2011,moody2013,neal2018}, we find that polarization has increased in both the US Senate and US House of Representatives, and that this polarization has largely mirrored partisan divisions along political party lines. We operationalized polarization using the level of a signed graph's structural balance, and therefore measure what \cite{neal2018} calls ``strong polarization,'' but have used two different measures of balance. We find that the two measures are highly correlated and both support the conclusion of increasing polarization.

The triangle index is easy to compute, but provides only a locally-aggregated measure of a graph's level of balance. In contrast, computing the frustration index is difficult, but it provides not only a global measure of a graph's level of balance, but also the optimal partitioning of vertices into internally cohesive but mutually antagonist groups. We have demonstrated a practical method for computing the exact value of frustration index and identifying the optimal partition in dense graphs of $|E|\gg50000$ that involves first obtaining upper and lower bounds, using exogenous node properties (e.g. legislators' political party affiliations), and solving a large-scale binary linear programming model. In the context of legislative networks, this method allows us to identify the most cohesive coalitions of legislators under conditions of balance theory.

Although our computational innovations make the identification of internally cohesive opposing coalitions practically feasible, we must also demonstrate that these coalitions are more informative than other simpler grouping possibilities. In the legislative context, we show that the partisan composition of these cohesive coalitions better explains the declining legislative effectiveness in the US House of Representatives than simply examining legislators' political party affiliations. This affirms Mayhew's claim that ``no theoretical treatment of the United States Congress that posits parties as analytic units will go very far'' \cite[p.27]{mayhew1974} but goes a step further by identifying an alternative analytic unit -- internally cohesive opposing coalitions -- that does have explanatory power. Importantly, coalitions appear useful only for explaining the legislative effectiveness of the House, but not the Senate. However, this is also consistent with existing political science theory that ``the lack of majority control of [procedural] processes in the Senate negates the possibility of significant party [or other group-based] effects in that body'' \cite[p.7]{monroe2008}. Therefore, in general terms, our empirical findings suggest that in legislative bodies where a sufficiently large group of legislators can influence procedural processes, the composition of the largest coalition is more important than the size of the majority party's majority. This is perhaps obvious in parliamentary systems where multi-party coalition forming is essential, but is noteworthy in the non-parliamentary US Congress.

These conclusions have some significant implications for both the future study of signed networks, and of the link between polarization and legislative effectiveness. First, by providing a practical method for computing the frustration index of relatively dense graphs, we hope to move the study of signed graphs beyond merely determining the level of balance, and toward the study of how the composition of mostly opposing groups impact other network dynamics. Second, our empirical findings suggest that research on polarization and its impact on the legislative process should look beyond political parties and partisanship to more subtle but influential forms of coordination, such as internally cohesive coalitions which are antagonist towards one another.

\section*{Materials and Methods}

Relations of collaboration and opposition between elected officials are difficult to collect directly because politicians have limited time to participate in surveys and have good reasons to conceal their true political relations. Therefore, studies of elected officials' political networks typically measure these relations indirectly, using bipartite projections focusing on their co-sponsorship of bills \cite{fowler_legislative_2006}, co-voting on bills \cite{moody2013,andris_2015,arinik_analysis_2018}, co-membership on committees \cite{porter_2005}, and co-attendance at press events \cite{desmarais_2015}. For a range of substantive reasons noted by \cite{neal2018} (e.g. relatively few bills are actually voted on, committee memberships are driven by such non-ideological factor such as seniority), we examine political relations from bill co-sponsorship.

Specifically, we use a signed network of inferred political relations among the members of the US House of Representatives, and among the members of the US Senate, in each session of Congress from 1979 to 2016 (96th session -- 114th session). The process for creating these signed networks is described in detail by \cite{neal2018} and they are available in a public \textit{Figshare} data repository \cite{Neal2019figshare}.

\subsection*{Inferring signed networks from co-sponsorship data}

\begin{figure}[ht]
    \centering
    \includegraphics[width=0.27\textwidth]{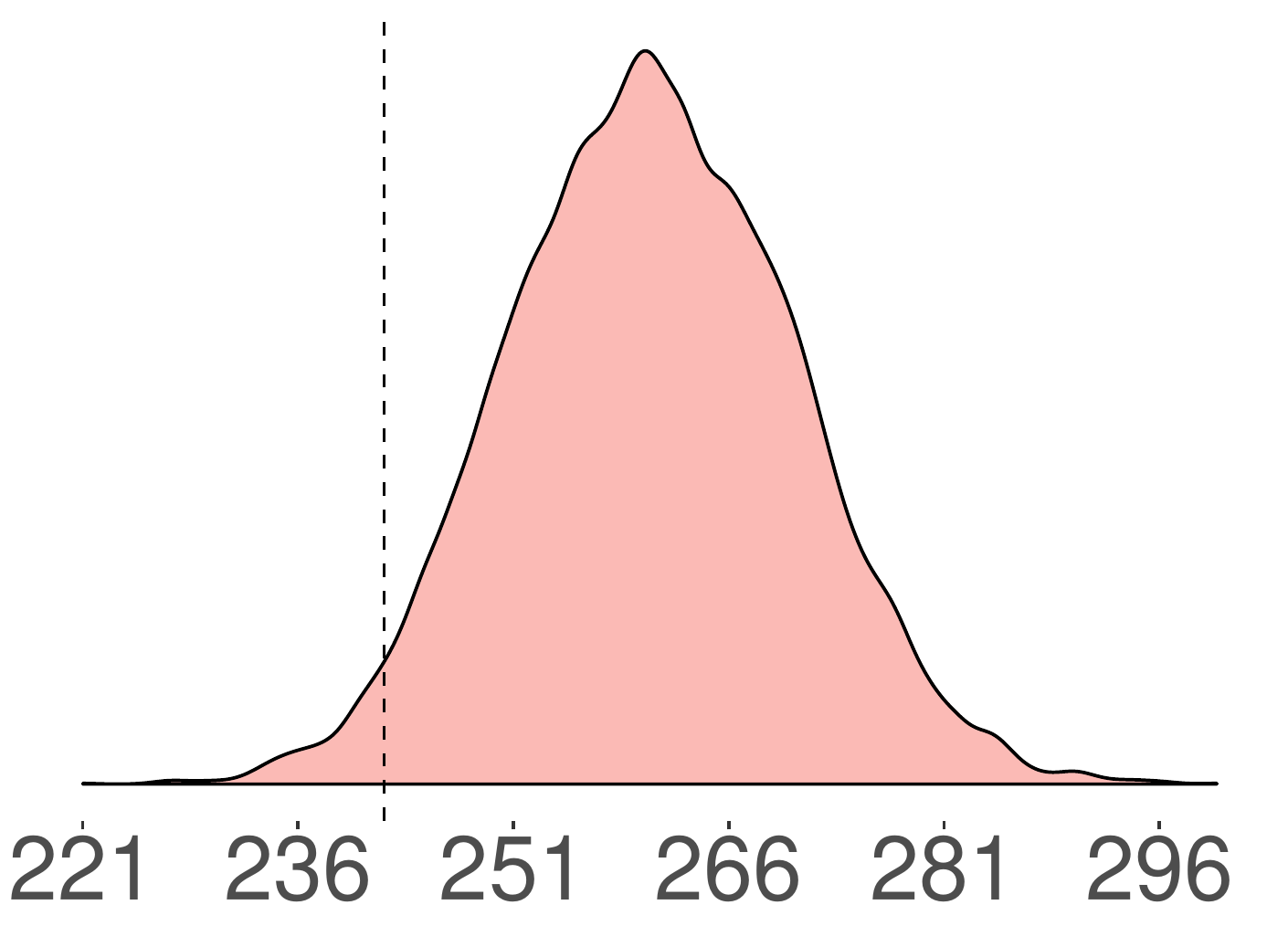}
    \caption{The observed number of co-sponsorships and the null model sampling distribution obtained using the SDSM for Rep. Earl Blumenhauer (D-OR3) and Rep. Sheila Jackson-Lee (D-TX18)}
    \label{fig:example}
\end{figure}

Importantly, all pairs of legislators co-sponsor at least some of the same bills, so we know that the mere existence of some co-sponsorships does not imply they collaborate, and that some number of co-sponsorships can actually indicate avoidance. In previous work \cite{neal2018}, a stochastic degree sequence model (SDSM) \cite{neal_backbone_2014} is used to define thresholds of significantly few and significantly many co-sponsorships by building the empirical sampling distribution of two legislators' joint co-sponsorships under a null model in which each legislator co-sponsored approximately the same number of bills and each bill received approximately the same number of co-sponsorships (i.e.\ holding approximately constant the legislator and bill degree sequence). To be more specific, given a bipartite graph $B$, Monte Carlo methods can be used to generate probability distributions $BB'_{ij}$ when $Pr(B_{ij}=1)$ is a function of the row and column marginals of $B$ \cite{neal2018}.  Decisions about whether a given dyad represents significantly few or significantly many co-sponsorships are made by comparing their observed number of joint co-sponsorships to the empirical sampling distribution using a two-tailed $\alpha = 0.05$ threshold.
For example, Fig. \ref{fig:example} 
shows that Rep. Earl Blumenhauer (D-OR3) and Rep. Sheila Jackson-Lee (D-TX18) were observed to have co-sponsored 242 of the same bills (dashed vertical line). The magnitude of joint co-sponsorships, and the fact that both representatives are Democrats, might lead one to conclude that they are collaborating. However, the shaded distribution shows the expected number of joint co-sponsorships under the SDSM null model in which each representative randomly chooses which bills to co-sponsor. Comparing these representatives' observed number of joint co-sponsorships to the null model expectation, we find that they co-sponsor significantly fewer of the same bills than would be expected at random and therefore define the edge between them as negative.

This approach differs from other methods of reducing weighted graphs to binary or signed graphs \cite{serrano2009,dianati2016} because it explicitly incorporates information from the original bipartite data (i.e.\ legislators linked to bills), thereby ensuring it is not lost when these data are projected as a unipartite graph. Additionally, \cite{neal2018} extracts signed backbone networks rather than the weighted bipartite projections because the weights in those projections are distorted by heterogeneity in the bipartite degree sequences (i.e.\ some legislators sponsor many bills, others sponsor few; \cite{latapy2008,neal_backbone_2014}).

Although data on earlier sessions are available, they were excluded because prior to the 96th session, House rules imposed a limit of 25 co-sponsors per bill, which artificially distorts co-sponsorship patterns and limits the usefulness of these data for inferring political networks \cite{thomas1993}. These data do not distinguish between a bill's ``sponsor'' and its ``co-sponsors'' because the former is simply the legislator whose name appears first in a potentially long list of legislators responsible for the bill's introduction.

Although these data represent a time-series of legislative interactions, we examine the networks cross-sectionally for two reasons. First, there are a large number of joiners and leavers in each new session as incumbents lose their seats, freshmen join Congress, or representatives become senators, making most dynamic models impractical to estimate. Second, although some political relationships develop over long periods of time, the effectiveness of any particular session of Congress can be evaluated independently.

\section*{Availability of data and code}
All the data and codes used in this study are publicly available with links and descriptions provided in the supplementary information.


\begin{thebibliography}{10}
	\urlstyle{rm}
	\expandafter\ifx\csname url\endcsname\relax
	\def\url#1{\texttt{#1}}\fi
	\expandafter\ifx\csname urlprefix\endcsname\relax\def\urlprefix{URL }\fi
	\expandafter\ifx\csname doiprefix\endcsname\relax\def\doiprefix{DOI: }\fi
	\providecommand{\bibinfo}[2]{#2}
	\providecommand{\eprint}[2][]{\url{#2}}
	
	\bibitem{robero2018}
	\bibinfo{author}{Ribeiro, H.~V.}, \bibinfo{author}{Alves, L. G.~A.},
	\bibinfo{author}{Martins, A.~F.}, \bibinfo{author}{Lenzi, E.~K.} \&
	\bibinfo{author}{Perc, M.}
	\newblock \bibinfo{journal}{\bibinfo{title}{{The dynamical structure of
				political corruption networks}}}.
	\newblock {\emph{\JournalTitle{Journal of Complex Networks}}}
	\textbf{\bibinfo{volume}{6}}, \bibinfo{pages}{989--1003},
	\doiprefix\url{10.1093/comnet/cny002} (\bibinfo{year}{2018}).
	
	\bibitem{colliri2019}
	\bibinfo{author}{Colliri, T.} \& \bibinfo{author}{Zhao, L.}
	\newblock \bibinfo{journal}{\bibinfo{title}{Analyzing the bills-voting dynamics
			and predicting corruption-convictions among {Brazilian} congressmen through
			temporal networks}}.
	\newblock {\emph{\JournalTitle{Scientific Reports}}}
	\textbf{\bibinfo{volume}{9}}, \bibinfo{pages}{16754},
	\doiprefix\url{10.1038/s41598-019-53252-9} (\bibinfo{year}{2019}).
	
	\bibitem{faustino2019}
	\bibinfo{author}{Faustino, J.}, \bibinfo{author}{Barbosa, H.},
	\bibinfo{author}{Ribeiro, E.} \& \bibinfo{author}{Menezes, R.}
	\newblock \bibinfo{journal}{\bibinfo{title}{{A data-driven network approach for
				characterization of political parties’ ideology dynamics}}}.
	\newblock {\emph{\JournalTitle{Applied Network Science}}}
	\textbf{\bibinfo{volume}{4}}, \bibinfo{pages}{48},
	\doiprefix\url{10.1007/s41109-019-0161-0} (\bibinfo{year}{2019}).
	
	\bibitem{Neal2019figshare}
	\bibinfo{author}{Neal, Z.}
	\newblock \bibinfo{title}{A Sign of the Times: Dataset of {US Congress} signed network backbones from co-sponsorship data, 1973-2016}.
	\newblock \bibinfo{howpublished}{\emph{figshare}
		\url{http://dx.doi.org/10.6084/m9.figshare.8096429}} (\bibinfo{year}{2019}).
	
	\bibitem{neal_backbone_2014}
	\bibinfo{author}{Neal, Z.}
	\newblock \bibinfo{journal}{\bibinfo{title}{The backbone of bipartite
			projections: {Inferring} relationships from co-authorship, co-sponsorship,
			co-attendance and other co-behaviors}}.
	\newblock {\emph{\JournalTitle{Social Networks}}}
	\textbf{\bibinfo{volume}{39}}, \bibinfo{pages}{84--97},
	\doiprefix\url{10.1016/j.socnet.2014.06.001} (\bibinfo{year}{2014}).
	
	\bibitem{neal2018}
	\bibinfo{author}{Neal, Z.}
	\newblock \bibinfo{journal}{\bibinfo{title}{A sign of the times? {Weak} and
			strong polarization in the {U.S.} {Congress}, 1973–2016}}.
	\newblock {\emph{\JournalTitle{Social Networks}}}
	\textbf{\bibinfo{volume}{60}}, \bibinfo{pages}{103 -- 112},
	\doiprefix\url{10.1016/j.socnet.2018.07.007} (\bibinfo{year}{2020}).
	
	\bibitem{terzi_spectral_2011}
	\bibinfo{author}{Terzi, E.} \& \bibinfo{author}{Winkler, M.}
	\newblock \bibinfo{title}{A spectral algorithm for computing social balance}.
	\newblock In \bibinfo{editor}{Frieze, A.}, \bibinfo{editor}{Horn, P.} \&
	\bibinfo{editor}{Pra{\l}at, P.} (eds.) \emph{\bibinfo{booktitle}{Proceedings
			of International Workshop on Algorithms and Models for the Web-Graph}}, WAW
	2011, \bibinfo{pages}{1--13}, \doiprefix\url{10.1007/978-3-642-21286-4_1}
	(\bibinfo{publisher}{Springer}, \bibinfo{year}{2011}).
	
	\bibitem{zaslavsky_balanced_1987}
	\bibinfo{author}{Zaslavsky, T.}
	\newblock \bibinfo{journal}{\bibinfo{title}{Balanced decompositions of a signed
			graph}}.
	\newblock {\emph{\JournalTitle{Journal of Combinatorial Theory, Series B}}}
	\textbf{\bibinfo{volume}{43}}, \bibinfo{pages}{1--13},
	\doiprefix\url{10.1016/0095-8956(87)90026-8} (\bibinfo{year}{1987}).
	
	\bibitem{facchetti_computing_2011}
	\bibinfo{author}{Facchetti, G.}, \bibinfo{author}{Iacono, G.} \&
	\bibinfo{author}{Altafini, C.}
	\newblock \bibinfo{journal}{\bibinfo{title}{Computing global structural balance
			in large-scale signed social networks}}.
	\newblock {\emph{\JournalTitle{Proceedings of the National Academy of
				Sciences}}} \textbf{\bibinfo{volume}{108}}, \bibinfo{pages}{20953--20958},
	\doiprefix\url{10.1073/pnas.1109521108} (\bibinfo{year}{2011}).
	
	\bibitem{aref2015measuring}
	\bibinfo{author}{Aref, S.} \& \bibinfo{author}{Wilson, M.~C.}
	\newblock \bibinfo{journal}{\bibinfo{title}{Measuring partial balance in signed
			networks}}.
	\newblock {\emph{\JournalTitle{Journal of Complex Networks}}}
	\textbf{\bibinfo{volume}{6}}, \bibinfo{pages}{566--595},
	\doiprefix\url{10.1093/comnet/cnx044} (\bibinfo{year}{2018}).
	
	\bibitem{harary_measurement_1959}
	\bibinfo{author}{Harary, F.}
	\newblock \bibinfo{journal}{\bibinfo{title}{On the measurement of structural
			balance}}.
	\newblock {\emph{\JournalTitle{Behavioral Science}}}
	\textbf{\bibinfo{volume}{4}}, \bibinfo{pages}{316--323},
	\doiprefix\url{10.1002/bs.3830040405} (\bibinfo{year}{1959}).
	
	\bibitem{harary_simple_1980}
	\bibinfo{author}{Harary, F.} \& \bibinfo{author}{Kabell, J.~A.}
	\newblock \bibinfo{journal}{\bibinfo{title}{A simple algorithm to detect
			balance in signed graphs}}.
	\newblock {\emph{\JournalTitle{Mathematical Social Sciences}}}
	\textbf{\bibinfo{volume}{1}}, \bibinfo{pages}{131--136},
	\doiprefix\url{10.1016/0165-4896(80)90010-4} (\bibinfo{year}{1980}).
	
	\bibitem{riker1962theory}
	\bibinfo{author}{Riker, W.~H.}
	\newblock \emph{\bibinfo{title}{The theory of political coalitions}}
	(\bibinfo{publisher}{Yale University Press}, \bibinfo{year}{1962}).
	
	\bibitem{huffner_separator-based_2010}
	\bibinfo{author}{H\"{u}ffner, F.}, \bibinfo{author}{Betzler, N.} \&
	\bibinfo{author}{Niedermeier, R.}
	\newblock \bibinfo{journal}{\bibinfo{title}{Separator-based data reduction for
			signed graph balancing}}.
	\newblock {\emph{\JournalTitle{Journal of Combinatorial Optimization}}}
	\textbf{\bibinfo{volume}{20}}, \bibinfo{pages}{335--360},
	\doiprefix\url{10.1007/s10878-009-9212-2} (\bibinfo{year}{2010}).
	
	\bibitem{gong_network_2017}
	\bibinfo{author}{Gong, M.}, \bibinfo{author}{Cai, Q.}, \bibinfo{author}{Ma,
		L.}, \bibinfo{author}{Wang, S.} \& \bibinfo{author}{Lei, Y.}
	\newblock \bibinfo{title}{Network structure balance analytics with evolutionary
		optimization}.
	\newblock In \emph{\bibinfo{booktitle}{Computational Intelligence for Network
			Structure Analytics}}, \bibinfo{pages}{135--199},
	\doiprefix\url{10.1007/978-981-10-4558-5_4} (\bibinfo{publisher}{Springer},
	\bibinfo{year}{2017}).
	
	\bibitem{traag_partitioning_2018}
	\bibinfo{author}{Traag, V.~A.}, \bibinfo{author}{Doreian, P.} \&
	\bibinfo{author}{Mrvar, A.}
	\newblock \bibinfo{title}{Partitioning signed networks}.
	\newblock In \bibinfo{editor}{Doreian, P.}, \bibinfo{editor}{Batagelj, V.} \&
	\bibinfo{editor}{Ferligoj, A.} (eds.) \emph{\bibinfo{booktitle}{Advances in
			network clustering and blockmodeling}}, chap.~\bibinfo{chapter}{8}
	(\bibinfo{publisher}{Wiley-Interscience}, \bibinfo{year}{2018}).
	
	\bibitem{hua_fast_2018}
	\bibinfo{author}{Hua, J.}, \bibinfo{author}{Yu, J.} \& \bibinfo{author}{Yang,
		M.-S.}
	\newblock \bibinfo{journal}{\bibinfo{title}{Fast clustering for signed graphs
			based on random walk gap}}.
	\newblock {\emph{\JournalTitle{Social Networks}}}
	\doiprefix\url{10.1016/j.socnet.2018.08.008} (\bibinfo{year}{2018}).
	
	\bibitem{brusco_partitioning_2019}
	\bibinfo{author}{Brusco, M.~J.} \& \bibinfo{author}{Doreian, P.}
	\newblock \bibinfo{journal}{\bibinfo{title}{Partitioning signed networks using
			relocation heuristics, tabu search, and variable neighborhood search}}.
	\newblock {\emph{\JournalTitle{Social Networks}}}
	\textbf{\bibinfo{volume}{56}}, \bibinfo{pages}{70--80},
	\doiprefix\url{10.1016/j.socnet.2018.08.007} (\bibinfo{year}{2019}).
	
	\bibitem{aref2016exact}
	\bibinfo{author}{Aref, S.}, \bibinfo{author}{Mason, A.~J.} \&
	\bibinfo{author}{Wilson, M.~C.}
	\newblock \bibinfo{journal}{\bibinfo{title}{A modeling and computational study
			of the frustration index in signed networks}}.
	\newblock {\emph{\JournalTitle{Networks}}} \textbf{\bibinfo{volume}{75}},
	\bibinfo{pages}{95--110}, \doiprefix\url{10.1002/net.21907}
	(\bibinfo{year}{2020}).
	
	\bibitem{heider_social_1944}
	\bibinfo{author}{Heider, F.}
	\newblock \bibinfo{journal}{\bibinfo{title}{Social perception and phenomenal
			causality}}.
	\newblock {\emph{\JournalTitle{Psychological Review}}}
	\textbf{\bibinfo{volume}{51}}, \bibinfo{pages}{358--378},
	\doiprefix\url{10.1037/h0055425} (\bibinfo{year}{1944}).
	
	\bibitem{cartwright_structural_1956}
	\bibinfo{author}{Cartwright, D.} \& \bibinfo{author}{Harary, F.}
	\newblock \bibinfo{journal}{\bibinfo{title}{Structural balance: a
			generalization of {Heider}'s theory}}.
	\newblock {\emph{\JournalTitle{Psychological Review}}}
	\textbf{\bibinfo{volume}{63}}, \bibinfo{pages}{277--293},
	\doiprefix\url{10.1037/h0046049} (\bibinfo{year}{1956}).
	
	\bibitem{aref2017computing}
	\bibinfo{author}{Aref, S.}, \bibinfo{author}{Mason, A.~J.} \&
	\bibinfo{author}{Wilson, M.~C.}
	\newblock \bibinfo{title}{Computing the line index of balance using integer
		programming optimisation}.
	\newblock In \bibinfo{editor}{Goldengorin, B.} (ed.)
	\emph{\bibinfo{booktitle}{Optimization Problems in Graph Theory}},
	\bibinfo{pages}{65--84}, \doiprefix\url{10.1007/978-3-319-94830-0_3}
	(\bibinfo{publisher}{Springer}, \bibinfo{year}{2018}).
	
	\bibitem{aref2017balance}
	\bibinfo{author}{Aref, S.} \& \bibinfo{author}{Wilson, M.~C.}
	\newblock \bibinfo{journal}{\bibinfo{title}{{Balance and frustration in signed
				networks}}}.
	\newblock {\emph{\JournalTitle{Journal of Complex Networks}}}
	\textbf{\bibinfo{volume}{7}}, \bibinfo{pages}{163--189},
	\doiprefix\url{10.1093/comnet/cny015} (\bibinfo{year}{2019}).
	
	\bibitem{harary1977graphing}
	\bibinfo{author}{Harary, F.}
	\newblock \bibinfo{journal}{\bibinfo{title}{Graphing conflict in international
			relations}}.
	\newblock {\emph{\JournalTitle{The papers of the Peace Science Society}}}
	\textbf{\bibinfo{volume}{27}}, \bibinfo{pages}{1--10} (\bibinfo{year}{1977}).
	
	\bibitem{abelson_symbolic_1958}
	\bibinfo{author}{Abelson, R.~P.} \& \bibinfo{author}{Rosenberg, M.~J.}
	\newblock \bibinfo{journal}{\bibinfo{title}{Symbolic psycho-logic: {A} model of
			attitudinal cognition}}.
	\newblock {\emph{\JournalTitle{Behavioral Science}}}
	\textbf{\bibinfo{volume}{3}}, \bibinfo{pages}{1--13},
	\doiprefix\url{10.1002/bs.3830030102} (\bibinfo{year}{1958}).
	
	\bibitem{KL1970}
	\bibinfo{author}{Kernighan, B.~W.} \& \bibinfo{author}{Lin, S.}
	\newblock \bibinfo{journal}{\bibinfo{title}{An efficient heuristic procedure
			for partitioning graphs}}.
	\newblock {\emph{\JournalTitle{Bell System Technical Journal}}}
	\textbf{\bibinfo{volume}{49}}, \bibinfo{pages}{291--307},
	\doiprefix\url{10.1002/j.1538-7305.1970.tb01770.x} (\bibinfo{year}{1970}).
	
	\bibitem{GN2002}
	\bibinfo{author}{Girvan, M.} \& \bibinfo{author}{Newman, M. E.~J.}
	\newblock \bibinfo{journal}{\bibinfo{title}{Community structure in social and
			biological networks}}.
	\newblock {\emph{\JournalTitle{Proceedings of the National Academy of
				Sciences}}} \textbf{\bibinfo{volume}{99}}, \bibinfo{pages}{7821--7826},
	\doiprefix\url{10.1073/pnas.122653799} (\bibinfo{year}{2002}).
	
	\bibitem{CNM2004}
	\bibinfo{author}{Clauset, A.}, \bibinfo{author}{Newman, M. E.~J.} \&
	\bibinfo{author}{Moore, C.}
	\newblock \bibinfo{journal}{\bibinfo{title}{Finding community structure in very
			large networks}}.
	\newblock {\emph{\JournalTitle{Phys. Rev. E}}} \textbf{\bibinfo{volume}{70}},
	\bibinfo{pages}{066111}, \doiprefix\url{10.1103/PhysRevE.70.066111}
	(\bibinfo{year}{2004}).
	
	\bibitem{Raghavan2007}
	\bibinfo{author}{Raghavan, U.~N.}, \bibinfo{author}{Albert, R.} \&
	\bibinfo{author}{Kumara, S.}
	\newblock \bibinfo{journal}{\bibinfo{title}{Near linear time algorithm to
			detect community structures in large-scale networks}}.
	\newblock {\emph{\JournalTitle{Phys. Rev. E}}} \textbf{\bibinfo{volume}{76}},
	\bibinfo{pages}{036106}, \doiprefix\url{10.1103/PhysRevE.76.036106}
	(\bibinfo{year}{2007}).
	
	\bibitem{Cordasco2010}
	\bibinfo{author}{{Cordasco}, G.} \& \bibinfo{author}{{Gargano}, L.}
	\newblock \bibinfo{title}{Community detection via semi-synchronous label
		propagation algorithms}.
	\newblock In \emph{\bibinfo{booktitle}{2010 IEEE International Workshop on:
			Business Applications of Social Network Analysis (BASNA)}},
	\bibinfo{pages}{1--8}, \doiprefix\url{10.1109/BASNA.2010.5730298}
	(\bibinfo{year}{2010}).
	
	\bibitem{Pares2017}
	\bibinfo{author}{Par{\'e}s, F.} \emph{et~al.}
	\newblock \bibinfo{title}{Fluid communities: A competitive, scalable and
		diverse community detection algorithm}.
	\newblock In \bibinfo{editor}{Cherifi, C.}, \bibinfo{editor}{Cherifi, H.},
	\bibinfo{editor}{Karsai, M.} \& \bibinfo{editor}{Musolesi, M.} (eds.)
	\emph{\bibinfo{booktitle}{Complex Networks {\&} Their Applications VI}},
	\bibinfo{pages}{229--240}, \doiprefix\url{10.1007/978-3-319-72150-7_19}
	(\bibinfo{publisher}{Springer International Publishing},
	\bibinfo{address}{Cham}, \bibinfo{year}{2018}).
	
	\bibitem{gurobi}
	\bibinfo{author}{{Gurobi Optimization Inc.}}
	\newblock \bibinfo{title}{Gurobi optimizer reference manual}
	(\bibinfo{year}{2018}).
	\newblock \bibinfo{note}{Url:
		\url{www.gurobi.com/documentation/8.0/refman/index.html} date accessed 1 June
		2018}.
	
	\bibitem{layman2006}
	\bibinfo{author}{Layman, G.~C.}, \bibinfo{author}{Carsey, T.~M.} \&
	\bibinfo{author}{Horowitz, J.~M.}
	\newblock \bibinfo{journal}{\bibinfo{title}{Party polarization in american
			politics: Characteristics, causes, and consequences}}.
	\newblock {\emph{\JournalTitle{Annual Review of Political Science}}}
	\textbf{\bibinfo{volume}{9}}, \bibinfo{pages}{83--110},
	\doiprefix\url{10.1146/annurev.polisci.9.070204.105138}
	(\bibinfo{year}{2006}).
	
	\bibitem{zhang2008}
	\bibinfo{author}{Zhang, Y.} \emph{et~al.}
	\newblock \bibinfo{journal}{\bibinfo{title}{Community structure in
			congressional cosponsorship networks}}.
	\newblock {\emph{\JournalTitle{Physica A}}} \textbf{\bibinfo{volume}{387}},
	\bibinfo{pages}{1705--1712}, \doiprefix\url{10.1016/j.physa.2007.11.004}
	(\bibinfo{year}{2008}).
	
	\bibitem{waugh2011}
	\bibinfo{author}{Waugh, A.~S.}, \bibinfo{author}{Pei, L.},
	\bibinfo{author}{Fowler, J.~H.}, \bibinfo{author}{Mucha, P.~J.} \&
	\bibinfo{author}{Porter, M.~A.}
	\newblock \bibinfo{journal}{\bibinfo{title}{Party polarization in congress: A
			network science approach}}.
	\newblock {\emph{\JournalTitle{arXiv}}}  (\bibinfo{year}{2011}).
	\newblock \bibinfo{note}{:0907.3509 (25 Jul 2011)}.
	
	\bibitem{moody2013}
	\bibinfo{author}{Moody, J.} \& \bibinfo{author}{Mucha, P.~J.}
	\newblock \bibinfo{journal}{\bibinfo{title}{Portrait of political party
			polarization}}.
	\newblock {\emph{\JournalTitle{Network Science}}} \textbf{\bibinfo{volume}{1}},
	\bibinfo{pages}{119--121}, \doiprefix\url{10.1017/nws.2012.3}
	(\bibinfo{year}{2013}).
	
	\bibitem{olson_measures_1972}
	\bibinfo{author}{Olson, D.~M.} \& \bibinfo{author}{Nonidez, C.~T.}
	\newblock \bibinfo{journal}{\bibinfo{title}{Measures of legislative performance
			in the {U.S.} {House} of {Representatives}}}.
	\newblock {\emph{\JournalTitle{Midwest Journal of Political Science}}}
	\textbf{\bibinfo{volume}{16}}, \bibinfo{pages}{269--277},
	\doiprefix\url{10.2307/2110060} (\bibinfo{year}{1972}).
	
	\bibitem{frantzich_who_1979}
	\bibinfo{author}{Frantzich, S.}
	\newblock \bibinfo{journal}{\bibinfo{title}{Who makes our laws? {The}
			legislative effectiveness of members of the {U.S.} congress}}.
	\newblock {\emph{\JournalTitle{Legislative Studies Quarterly}}}
	\textbf{\bibinfo{volume}{4}}, \bibinfo{pages}{409--428},
	\doiprefix\url{10.2307/439582} (\bibinfo{year}{1979}).
	
	\bibitem{volden2014}
	\bibinfo{author}{Volden, C.} \& \bibinfo{author}{Wiseman, A.~E.}
	\newblock \emph{\bibinfo{title}{Legislative effectiveness in the United States
			Congress: {The} lawmakers}} (\bibinfo{publisher}{Cambridge university press},
	\bibinfo{year}{2014}).
	
	\bibitem{mayhew2005}
	\bibinfo{author}{Mayhew, D.~R.}
	\newblock \emph{\bibinfo{title}{Divided we govern: Party control, lawmaking,
			and investigations, 1946-2002}} (\bibinfo{publisher}{Yale university press},
	\bibinfo{year}{2005}).
	
	\bibitem{moore_legislative_1969}
	\bibinfo{author}{Moore, D.~W.}
	\newblock \bibinfo{journal}{\bibinfo{title}{Legislative effectiveness and
			majority party size: A test in the indiana house}}.
	\newblock {\emph{\JournalTitle{The Journal of Politics}}}
	\textbf{\bibinfo{volume}{31}}, \bibinfo{pages}{1063--1079},
	\doiprefix\url{10.2307/2128358} (\bibinfo{year}{1969}).
	
	\bibitem{mayhew1974}
	\bibinfo{author}{Mayhew, D.~R.}
	\newblock \emph{\bibinfo{title}{Congress: The Electoral Connection}}
	(\bibinfo{publisher}{Yale university press}, \bibinfo{year}{1974}).
	
	\bibitem{monroe2008}
	\bibinfo{author}{Monroe, N.~W.}, \bibinfo{author}{Roberts, J.~M.} \&
	\bibinfo{author}{Rohde, D.~W.}
	\newblock \emph{\bibinfo{title}{Why Not Parties? Party Effects in the United
			States Senate}} (\bibinfo{publisher}{University of Chicago Press},
	\bibinfo{year}{2008}).
	
	\bibitem{fowler_legislative_2006}
	\bibinfo{author}{Fowler, J.~H.}
	\newblock \bibinfo{journal}{\bibinfo{title}{Legislative cosponsorship networks
			in the {US} {House} and {Senate}}}.
	\newblock {\emph{\JournalTitle{Social Networks}}}
	\textbf{\bibinfo{volume}{28}}, \bibinfo{pages}{454--465},
	\doiprefix\url{10.1016/j.socnet.2005.11.003} (\bibinfo{year}{2006}).
	
	\bibitem{andris_2015}
	\bibinfo{author}{Andris, C.} \emph{et~al.}
	\newblock \bibinfo{journal}{\bibinfo{title}{The rise of partisanship and
			super-cooperators in the {U.S.} {House} of {Representatives}}}.
	\newblock {\emph{\JournalTitle{PloS one}}} \textbf{\bibinfo{volume}{10}},
	\bibinfo{pages}{1--14}, \doiprefix\url{10.1371/journal.pone.0123507}
	(\bibinfo{year}{2015}).
	
	\bibitem{arinik_analysis_2018}
	\bibinfo{author}{Arinik, N.}, \bibinfo{author}{Figueiredo, R.} \&
	\bibinfo{author}{Labatut, V.}
	\newblock \bibinfo{journal}{\bibinfo{title}{Analysis of roll-calls in the
			{European} parliament by multiple partitioning of multiplex signed
			networks}}.
	\newblock {\emph{\JournalTitle{Social Networks {(in press)}}}}
	(\bibinfo{year}{2019}).
	\newblock \bibinfo{note}{Doi:
		\href{http://doi.org/10.1016/j.socnet.2019.02.001}{10.1016/j.socnet.2019.02.001}
		(26 November 2018)}.
	
	\bibitem{porter_2005}
	\bibinfo{author}{Porter, M.~A.}, \bibinfo{author}{Mucha, P.~J.},
	\bibinfo{author}{Newman, M. E.~J.} \& \bibinfo{author}{Warmbrand, C.~M.}
	\newblock \bibinfo{journal}{\bibinfo{title}{A network analysis of committees in
			the {U.S.} {House} of {Representatives}}}.
	\newblock {\emph{\JournalTitle{Proceedings of the National Academy of
				Sciences}}} \textbf{\bibinfo{volume}{102}}, \bibinfo{pages}{7057--7062},
	\doiprefix\url{10.1073/pnas.0500191102} (\bibinfo{year}{2005}).
	
	\bibitem{desmarais_2015}
	\bibinfo{author}{Desmarais, B.~A.}, \bibinfo{author}{Moscardelli, V.~G.},
	\bibinfo{author}{Schaffner, B.~F.} \& \bibinfo{author}{Kowal, M.~S.}
	\newblock \bibinfo{journal}{\bibinfo{title}{Measuring legislative
			collaboration: The {Senate} press events network}}.
	\newblock {\emph{\JournalTitle{Social Networks}}}
	\textbf{\bibinfo{volume}{40}}, \bibinfo{pages}{43--54},
	\doiprefix\url{10.1016/j.socnet.2014.07.006} (\bibinfo{year}{2015}).
	
	\bibitem{serrano2009}
	\bibinfo{author}{Serrano, M.~{\'A}.}, \bibinfo{author}{Bogu{\~n}{\'a}, M.} \&
	\bibinfo{author}{Vespignani, A.}
	\newblock \bibinfo{journal}{\bibinfo{title}{Extracting the multiscale backbone
			of complex weighted networks}}.
	\newblock {\emph{\JournalTitle{Proceedings of the National Academy of
				Sciences}}} \textbf{\bibinfo{volume}{106}}, \bibinfo{pages}{6483--6488},
	\doiprefix\url{10.1073/pnas.0808904106} (\bibinfo{year}{2009}).
	
	\bibitem{dianati2016}
	\bibinfo{author}{Dianati, N.}
	\newblock \bibinfo{journal}{\bibinfo{title}{Unwinding the hairball graph:
			Pruning algorithms for weighted complex networks}}.
	\newblock {\emph{\JournalTitle{Phys. Rev. E}}} \textbf{\bibinfo{volume}{93}},
	\bibinfo{pages}{012304}, \doiprefix\url{10.1103/PhysRevE.93.012304}
	(\bibinfo{year}{2016}).
	
	\bibitem{latapy2008}
	\bibinfo{author}{Latapy, M.}, \bibinfo{author}{Magnien, C.} \&
	\bibinfo{author}{Vecchio, N.~D.}
	\newblock \bibinfo{journal}{\bibinfo{title}{Basic notions for the analysis of
			large two-mode networks}}.
	\newblock {\emph{\JournalTitle{Social Networks}}}
	\textbf{\bibinfo{volume}{30}}, \bibinfo{pages}{31 -- 48},
	\doiprefix\url{10.1016/j.socnet.2007.04.006} (\bibinfo{year}{2008}).
	
	\bibitem{thomas1993}
	\bibinfo{author}{Thomas, S.} \& \bibinfo{author}{Grofman, B.}
	\newblock \bibinfo{journal}{\bibinfo{title}{The effects of congressional rules
			about bill cosponsorship on duplicate bills: changing incentives for credit
			claiming}}.
	\newblock {\emph{\JournalTitle{Public Choice}}} \textbf{\bibinfo{volume}{75}},
	\bibinfo{pages}{93--98}, \doiprefix\url{10.1007/BF01053883}
	(\bibinfo{year}{1993}).
	
	\bibitem{olzak_friends_2016}
	\bibinfo{author}{Olzak, S.}, \bibinfo{author}{Soule, S.~A.},
	\bibinfo{author}{Coddou, M.} \& \bibinfo{author}{Muñoz, J.}
	\newblock \bibinfo{journal}{\bibinfo{title}{Friends or foes? {How} social
			movement allies affect the passage of legislation in the {U.S.} {Congress}}}.
	\newblock {\emph{\JournalTitle{Mobilization: An International Quarterly}}}
	\textbf{\bibinfo{volume}{21}}, \bibinfo{pages}{213--230},
	\doiprefix\url{10.17813/1086-671X-21-2-213} (\bibinfo{year}{2016}).
	\newblock \eprint{https://doi.org/10.17813/1086-671X-21-2-213}.
	
	\bibitem{anderson_keys_2003}
	\bibinfo{author}{Anderson, W.~D.}, \bibinfo{author}{Box-Steffensmeier, J.~M.}
	\& \bibinfo{author}{Sinclair-Chapman, V.}
	\newblock \bibinfo{journal}{\bibinfo{title}{The keys to legislative success in
			the {U.S.} {House} of {Representatives}}}.
	\newblock {\emph{\JournalTitle{Legislative Studies Quarterly}}}
	\textbf{\bibinfo{volume}{28}}, \bibinfo{pages}{357--386},
	\doiprefix\url{10.3162/036298003X200926} (\bibinfo{year}{2003}).
	
	\bibitem{finocchiaro_war_2008}
	\bibinfo{author}{Finocchiaro, C.~J.} \& \bibinfo{author}{Rohde, D.~W.}
	\newblock \bibinfo{journal}{\bibinfo{title}{War for the floor: Partisan theory
			and agenda control in the {U.S.} {House} of {Representatives}}}.
	\newblock {\emph{\JournalTitle{Legislative Studies Quarterly}}}
	\textbf{\bibinfo{volume}{33}}, \bibinfo{pages}{35--61},
	\doiprefix\url{10.3162/036298008783743273} (\bibinfo{year}{2008}).
	
	\bibitem{poole1984polarization}
	\bibinfo{author}{Poole, K.~T.} \& \bibinfo{author}{Rosenthal, H.}
	\newblock \bibinfo{journal}{\bibinfo{title}{The polarization of american
			politics}}.
	\newblock {\emph{\JournalTitle{The Journal of Politics}}}
	\textbf{\bibinfo{volume}{46}}, \bibinfo{pages}{1061--1079},
	\doiprefix\url{10.2307/2131242} (\bibinfo{year}{1984}).
	
	\bibitem{poole2000congress}
	\bibinfo{author}{Poole, K.~T.} \& \bibinfo{author}{Rosenthal, H.}
	\newblock \emph{\bibinfo{title}{Congress: A political-economic history of roll
			call voting}} (\bibinfo{publisher}{Oxford University Press on Demand},
	\bibinfo{year}{2000}).
	
	\bibitem{cox2002measuring}
	\bibinfo{author}{Cox, G.~W.} \& \bibinfo{author}{Poole, K.~T.}
	\newblock \bibinfo{journal}{\bibinfo{title}{On measuring partisanship in
			roll-call voting: The {US House} of {Representatives}, 1877-1999}}.
	\newblock {\emph{\JournalTitle{American Journal of Political Science}}}
	\bibinfo{pages}{477--489}, \doiprefix\url{10.2307/3088393}
	(\bibinfo{year}{2002}).
	
\end{thebibliography}

\section*{Author contributions}
S.A. developed and implemented new methods and algorithms, obtained the results, and prepared supplementary information; Z.N. prepared network data, ran statistical models, and analyzed the results; both authors contributed to designing and conducting the research and writing the paper.

\section*{Acknowledgments}
The authors acknowledge Center for eResearch at University of Auckland for providing access to high-performance computers. There is no funding to be reported for this study. 


\section*{Additional Information}
The authors declare neither financial nor non-financial competing interests.

\section*{Supplementary information for \\
Detecting coalitions by optimally partitioning signed networks of political collaboration}

This document describes all materials and methods for the article “Detecting coalitions by optimally partitioning signed networks of political collaboration” by Samin Aref and Zachary Neal.\\

\noindent
\textbf{Supplementary Information:}

\noindent
Supplementary Text\\
Figs. \ref{fig-low-low} to \ref{fig:largest}\\
Tables \ref{tab:senatenetworks} to \ref{tab:housetable}\\
Captions for Movies S1 to S2\\
Captions for Databases S1 to S2\\
References \textit{(53-58)}\\
\\
\textbf{Other Supplementary Materials for this manuscript include the following:} 

\noindent
Movies S1 to S2\\
Databases S1 to S2\\

\section*{Supplementary Text}

\subsection*{Data availability} 
All network data and numerical results related to this study are publicly available with links provided in this document. The code for the optimization models used is this study is publicly available on a Github repository.

\subsection*{Analyzing networks of legislators}

Most research on performance of political systems, and on the link between polarization and legislative effectiveness, has focused on legislators' ideological positions \cite{olzak_friends_2016}, role of political parties \cite{anderson_keys_2003,finocchiaro_war_2008}, and majority party size \cite{moore_legislative_1969} within legislative chambers. However, others have suggested that a focus on parties to explain the dynamics of the US Congress is misguided \cite{mayhew1974}. Parties are administrative conveniences that facilitate coordination and often serve as a useful heuristic for their members' ideology, but because party affiliation is different from ideological position, a focus on political parties oversimplifies matters by assuming within-party ideological homogeneity and between-party ideological heterogeneity. Therefore, in this paper, we adopt a different approach, focusing more on networks of collaborations between legislators during a two-year session and less on legislators' political party affiliations. We find that this approach -- examining polarization from the perspective of networks and structural balance -- offers a better explanation of legislative effectiveness than political parties.

Our method of analysis is different from the conventional methods of indexing legislators partisanship \cite{poole1984polarization,poole2000congress,cox2002measuring} which place each legislator on a scale of liberal to conservative. These methods are shown to produce results correlating with important historical events in the US politics and therefore are standard practice in quantifying polarization \cite{andris_2015}. While these methods indicate the political climate as a whole, they are not designed to take network relations of legislators into account.

\subsection*{Statistical analysis}
The models shown in Fig. 5 (of the article) and discussed in the ``Mediation in bill passage'' section were estimated using Stata SE 13.1. All models were estimated separately for the US House of Representative and US Senate, and only standardized coefficients are reported. Model A was estimated as an ordinary bivariate linear regression using the following code in \textit{Stata} software (release 13).
\begin{verbatim}
reg rate session, beta    
\end{verbatim}

Models B and C were estimated as structural equation models with maximum likelihood estimation using the following codes.

\begin{verbatim}
sem (party <- session) (rate <- party session)
sem (coalition <- session) (rate <- coalition session)
\end{verbatim}

This estimation approach allows us to explicitly estimate the total indirect effect of time mediated by party control (in model B) or coalition partisanship (model C), and thus to test whether these variables help explain observed declines in bill passage rates.


\subsection*{Solving the continuous and discrete optimization models}

The proposed continuous optimization model can be solved by any mathematical programming solver which supports linear programming (LP) models. In the Github repository, we explain using Gurobi solver (version 8.0) for solving the proposed LP model. The proposed discrete optimization model can be solved by any mathematical programming solver which supports 0/1 linear programming (binary linear) models. In the Github repository, we share our code for using Gurobi solver (version 8.0) to solve the proposed binary linear model.

The code for both the continuous and the discrete optimization models is available on a Github repository at \url{https://github.com/saref/frustration-index-dense}.

\subsection*{Using Gurobi for solving mathematical programming models}

Our proposed algorithms are developed in Python 3.7 based on the mathematical programming models in \cite{aref2017computing,aref2016exact} for computing the frustration index.

These optimization algorithms are distributed under an Attribution-NonCommercial-ShareAlike 4.0 International (CC BY-NC-SA 4.0) license. This means that one can use these algorithms for non-commercial purposes provided that they provide proper attribution for them by citing \cite{aref2017computing,aref2016exact} and the current article. Copies or adaptations of the algorithms should be released under the similar license.

The following steps outline the process for academics to install the required software (\textit{Gurobi} solver \cite{gurobi}) on your computer to be able to run the optimization algorithms:

\begin{enumerate}
	\item 
	Download and install Anaconda (Python 3.7 version) which allows you to run a Jupyter code. It can be downloaded from \url{https://www.anaconda.com/distribution/}. Note that you must select your operating system first and download the corresponding installer.
	
	\item 
	Register for an account on \url{gurobi.com/registration-general-reg/} to get a free academic license for using Gurobi. Note that Gurobi is a commercial software, but it can be registered with a free academic license if the user is affiliated with a recognized degree-granting academic institution. This involves creating an account on Gurobi website to be able to request a free academic license in step 5.
	
	\item 
	Download and install Gurobi Optimizer (versions 8.0 and above are recommended) which can be downloaded from \url{https://www.gurobi.com/downloads/gurobi-optimizer-eula/} after reading and agreeing to Gurobi's End User License Agreement.
	
	\item
	Install Gurobi into Anaconda. You do this by first adding the Gurobi channel to your Anaconda channels and then installing the Gurobi package from this channel.
	
	From a terminal window issue the following command to add the Gurobi channel to your default search list
	
	\begin{verbatim}
	conda config --add channels http://conda.anaconda.org/gurobi
	\end{verbatim} 
	
	Now issue the following command to install the Gurobi package
	
	\begin{verbatim}
	conda install gurobi
	\end{verbatim}
	
	\item 
	Request an academic license from \url{gurobi.com/downloads/end-user-license-agreement-academic/} and install the license on your computer following the instructions given on Gurobi license page.
	
	Completing these steps is explained in the following links (for version 8.1):
	
	for Windows \url{https://www.gurobi.com/documentation/8.1/quickstart_windows/installing_the_anaconda_py.html},
	
	for Linux \url{gurobi.com/documentation/8.1/quickstart_linux/installing_the_anaconda_py.html}, and
	
	for Mac \url{gurobi.com/documentation/8.1/quickstart_mac/installing_the_anaconda_py.html}.
	
	After following the instructions above, open Jupyter Notebook which takes you to an environment (a new tab on your browser pops up on your screen) where you can open the main code (which is a file with .ipynb extension).
	
\end{enumerate}

\begin{figure}[ht]
	\centering
	\includegraphics[width=\textwidth]{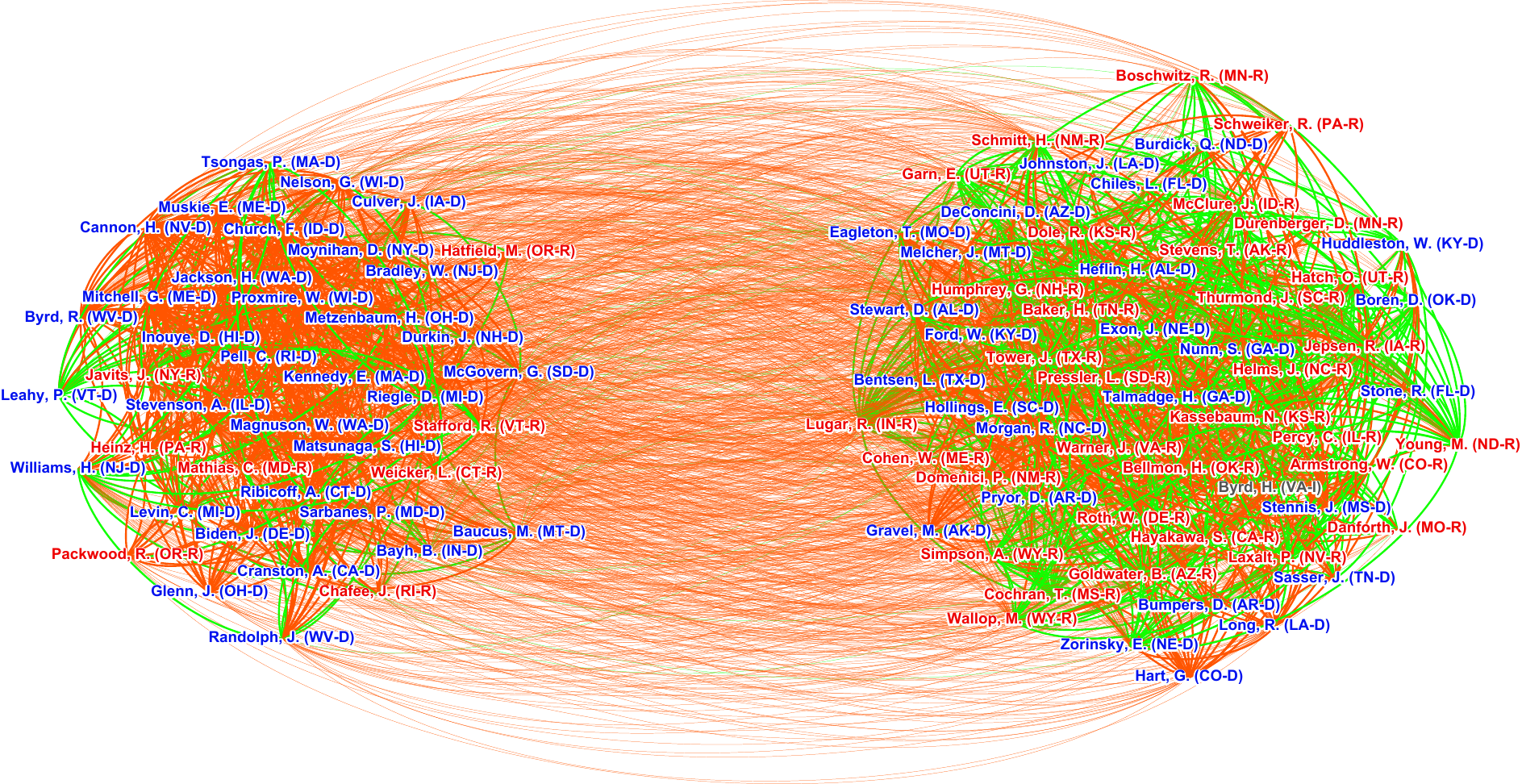}
	\caption{Opposing coalitions in the 96th session of the US Senate (1979). This network shows low balance and low coalition control.}
	\label{fig-low-low}
\end{figure}

\begin{figure}
	\centering
	\includegraphics[width=\textwidth]{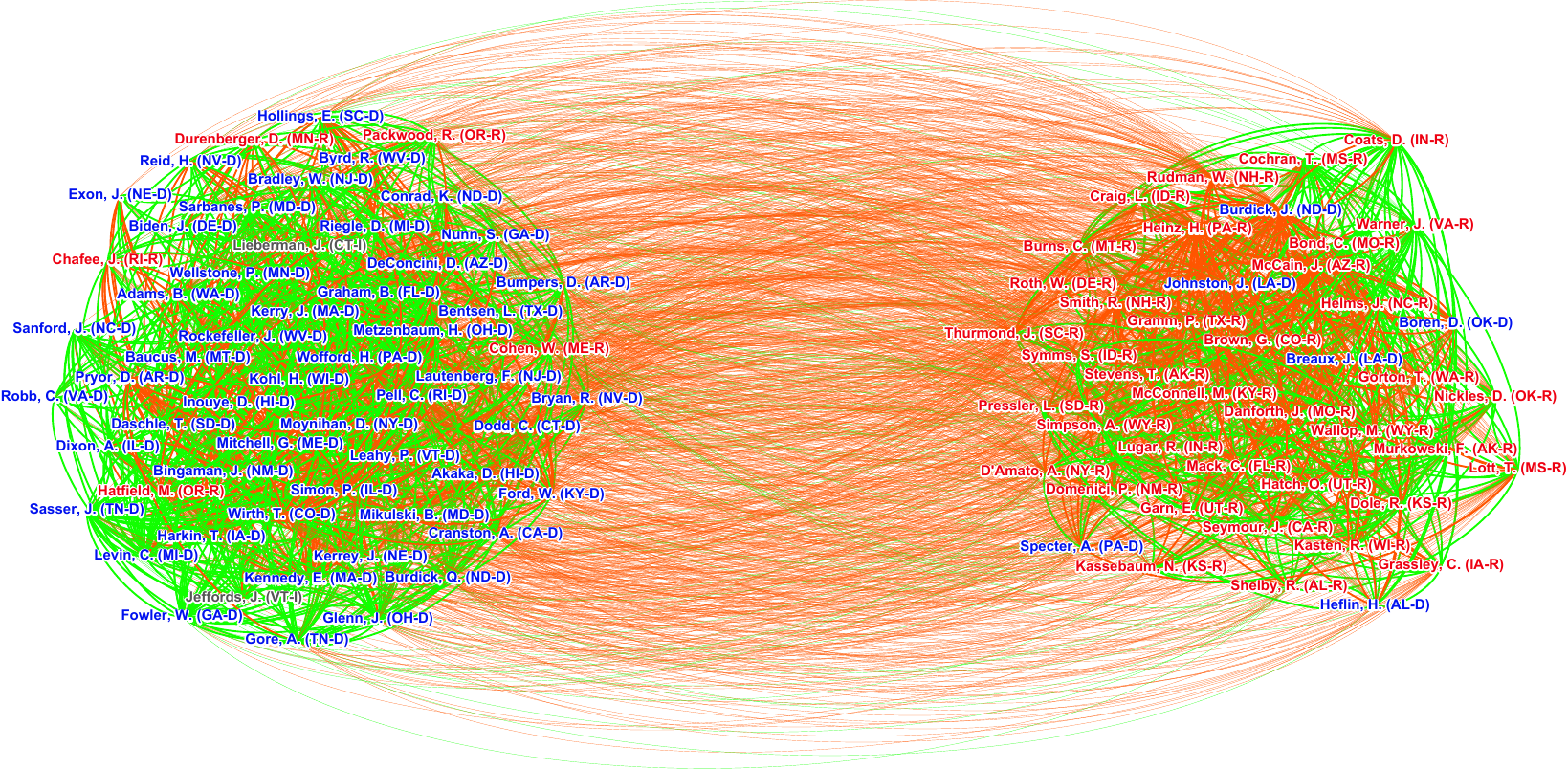}
	\caption{Opposing coalitions in the 102th session of the US Senate (1991). This network shows low balance and high coalition control.}
	\label{fig-low-high}
\end{figure}

\begin{figure}
	\centering
	\includegraphics[width=\textwidth]{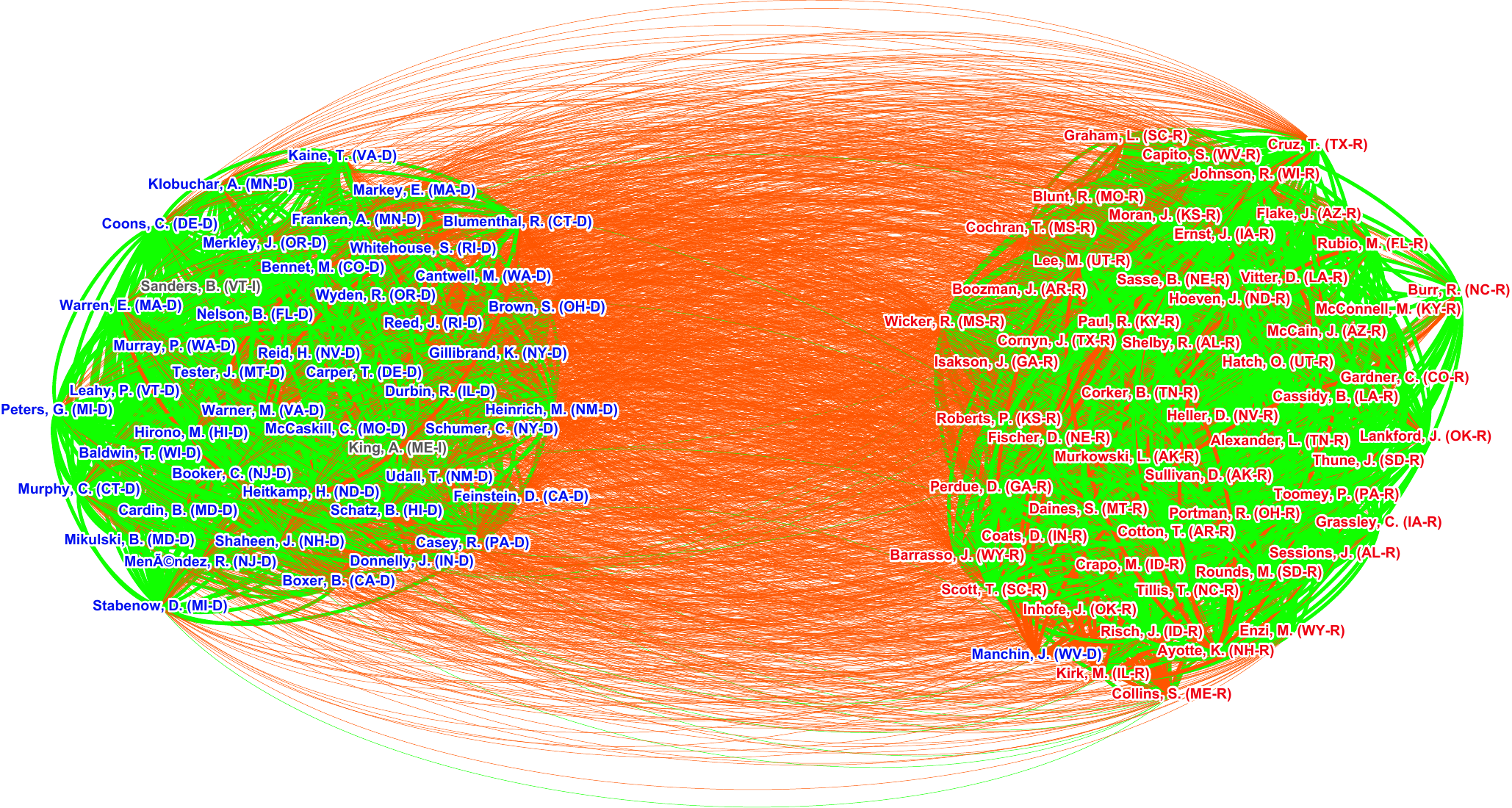}
	\caption{Opposing coalitions in the 114th session of the US Senate (2015). This network shows high balance and high coalition control.}
	\label{fig-high-high}
\end{figure}

\subsection*{Visualization of opposing coalitions in Senate networks (Figures \ref{fig-low-low} to \ref{fig-high-high})}

Figs. \ref{fig-low-low}--\ref{fig-high-high} show the opposing coalitions in three selected Senate networks. Green and orange edges represent significantly many and significantly few co-sponsorships respectively. Blue and red nodes represent Democratic- and Republican-affiliated legislators respectively and nodes for independent legislators are shown in gray.

In Fig. \ref{fig-low-low}, the network has a low level of balance $F(G)=0.524$ (low polarization) and the larger coalition is relatively heterogeneous, which is reflected in the relatively small value of coalition control ($0.559$). In Fig. \ref{fig-low-high}, the network has a low level of balance $F(G)=0.603$ (low polarization) and the larger coalition is relatively homogeneous, which is reflected in the relatively high value of coalition control ($0.911$). In Fig. \ref{fig-high-high}, the network has a high level of balance $F(G)=0.953$ (high polarization) and the larger coalition is homogeneous (coalition control equals $0.981$). The level of balance (coalition control) can also be observed from the colors of the edges (nodes) in Figs. \ref{fig-low-low}--\ref{fig-high-high}.


\clearpage

\subsection*{Size of controlling coalitions}

Using the optimal values of the $x_i$ variables obtained by solving the discrete optimization model, we partition nodes of each network into two groups (subsets $X^*$, $V \setminus X^*$), namely nodes associated with $x_i$ variables taking value $0$ in the optimal solution and nodes whose corresponding variables take value $1$ in the optimal solution.

\begin{figure}[ht]
	\centering
		\includegraphics[width=1\textwidth]{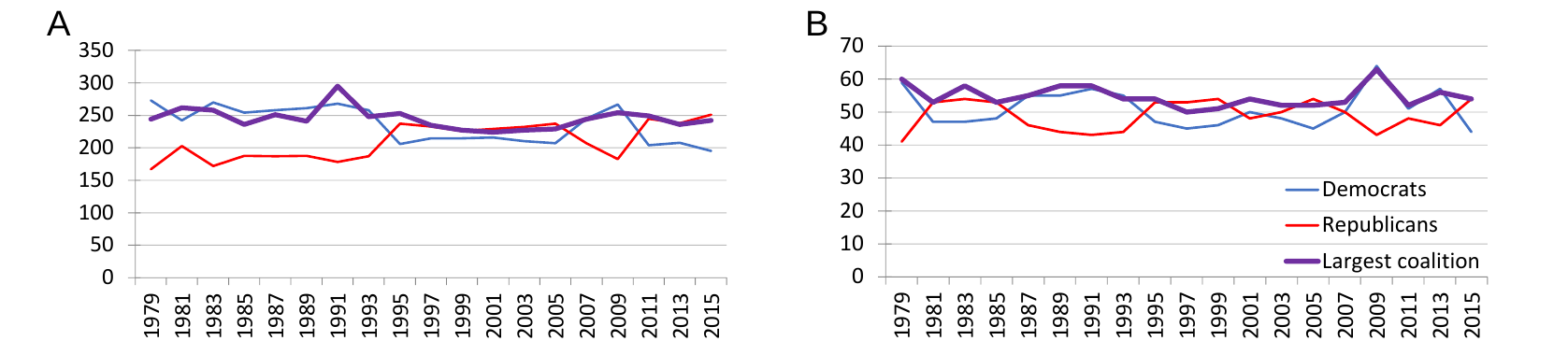}
	\caption{The number of legislators from the two main parties and the size of the largest coalition in (A) US House of Representatives and (B) US Senate over the time period 1979-2016}
	\label{fig:largest}
\end{figure}

For each signed network, either $X^*$ or $V \setminus X^*$ has the larger set cardinality and therefore represents the largest coalition for the corresponding session. Fig. \ref{fig:largest} 

shows the size of the largest and therefore controlling coalitions (winning coalitions \cite{riker1962theory}) in each signed network alongside the number of Democrats and Republicans in each session for both chambers. 

The controlling coalitions that our partitioning algorithm produces seem to be consistent with Riker's theory of \textit{minimum winning coalitions} \cite{riker1962theory} which argues that politicians try to form winning (larger than majority limit) coalitions which are cohesive and minimal in size.   

\subsection*{Legislative effectiveness}
We compute our central outcome, legislative effectiveness, as the fraction of bills introduced in a chamber that are eventually signed into law by the president. Tables \ref{tab:senatetable} and \ref{tab:housetable} 
report the number of bills introduced in the respective chamber and the number of these bills signed into law, as well as the passage rate, which measures legislative effectiveness. As the values in Tables \ref{tab:senatetable} and \ref{tab:housetable} illustrate, the passage rate has generally declined over time. Importantly, this is not simply due to the introduction of more frivolous bills and thus is not a simple function of an increase in the denominator (i.e.\ bills introduced), which has fluctuated around a mean of 3386 in the Senate and 6483 in the House.

\subsection*{Additional numerical results for the Senate and House networks (Tables \ref{tab:senatenetworks} to \ref{tab:housetable})}

\begin{sidewaystable}\centering
	\caption{Detailed properties and results for Senate networks}
	\begin{tabular}{lllllllllll}
		\hline
		Session & Year & $n$ & $m$  & density & $m^-$ & $m^+$ & $T(G)$ & $F(G)$ & $L(G)$ & $Y^*$ \\ \hline
		96      & 1979 & 101 & 2275 & 0.450   & 1870  & 405   & 0.410  & 0.524  & 541    & 541   \\
		97      & 1981 & 101 & 2073 & 0.410   & 1639  & 434   & 0.432  & 0.580  & 435    & 435   \\
		98      & 1983 & 101 & 2194 & 0.434   & 1676  & 518   & 0.511  & 0.521  & 525    & 525   \\
		99      & 1985 & 101 & 2177 & 0.431   & 1642  & 535   & 0.566  & 0.586  & 451    & 451   \\
		100     & 1987 & 101 & 2143 & 0.424   & 1535  & 608   & 0.594  & 0.525  & 509    & 509   \\
		101     & 1989 & 101 & 2445 & 0.484   & 1666  & 779   & 0.606  & 0.551  & 549    & 549   \\
		102     & 1991 & 102 & 2479 & 0.481   & 1768  & 711   & 0.635  & 0.603  & 492    & 492   \\
		103     & 1993 & 101 & 2257 & 0.447   & 1633  & 624   & 0.728  & 0.691  & 349    & 349   \\
		104     & 1995 & 102 & 2324 & 0.451   & 1715  & 609   & 0.777  & 0.811  & 220    & 220   \\
		105     & 1997 & 100 & 3002 & 0.606   & 2112  & 890   & 0.821  & 0.839  & 241    & 241   \\
		106     & 1999 & 102 & 2930 & 0.569   & 2108  & 822   & 0.852  & 0.816  & 269    & 269   \\
		107     & 2001 & 101 & 2522 & 0.499   & 1844  & 678   & 0.859  & 0.735  & 334    & 334   \\
		108     & 2003 & 100 & 2387 & 0.482   & 1759  & 628   & 0.862  & 0.828  & 205    & 205   \\
		109     & 2005 & 101 & 2823 & 0.559   & 2048  & 775   & 0.848  & 0.834  & 235    & 235   \\
		110     & 2007 & 102 & 2779 & 0.540   & 1934  & 845   & 0.853  & 0.851  & 207    & 207   \\
		111     & 2009 & 109 & 3645 & 0.619   & 2692  & 953   & 0.806  & 0.710  & 528    & 528   \\
		112     & 2011 & 101 & 3914 & 0.775   & 2693  & 1221  & 0.847  & 0.786  & 418    & 418   \\
		113     & 2013 & 105 & 3932 & 0.720   & 2554  & 1378  & 0.890  & 0.877  & 241    & 241   \\
		114     & 2015 & 100 & 3696 & 0.747   & 2261  & 1435  & 0.884  & 0.953  & 86     & 86    \\ \hline
	\end{tabular}
	\label{tab:senatenetworks}
\end{sidewaystable}

\begin{sidewaystable}\centering
	\caption{Democrats (Dems), Republicans (Reps), controlling coalitions (CC), and bill passage in the Senate}
	\label{tab:senatetable}
	\begin{tabular}{lllllllllll}
		\hline
		Session & Dems & Reps & Size of CC & Dems in CC & Reps in CC & Party control & Coalition control & Bills introduced & Signed into law & Passage rate \\ \hline
		96      & 59   & 41   & 60         & 26         & 33         & 18            & 0.559             & 3480             & 257             & 0.074        \\
		97      & 47   & 53   & 53         & 39         & 14         & 6             & 0.736             & 3396             & 210             & 0.062        \\
		98      & 47   & 54   & 58         & 44         & 14         & 7             & 0.759             & 3455             & 283             & 0.082        \\
		99      & 48   & 53   & 53         & 41         & 12         & 5             & 0.774             & 3386             & 304             & 0.090        \\
		100     & 55   & 46   & 55         & 16         & 39         & 9             & 0.709             & 3319             & 299             & 0.090        \\
		101     & 55   & 44   & 58         & 18         & 40         & 11            & 0.690             & 3659             & 277             & 0.076        \\
		102     & 57   & 43   & 58         & 51         & 5          & 14            & 0.911             & 3736             & 198             & 0.053        \\
		103     & 55   & 44   & 54         & 49         & 3          & 11            & 0.942             & 2801             & 172             & 0.061        \\
		104     & 47   & 53   & 54         & 4          & 50         & 6             & 0.926             & 2264             & 81              & 0.036        \\
		105     & 45   & 53   & 50         & 0          & 50         & 8             & 1.000             & 2715             & 141             & 0.052        \\
		106     & 46   & 54   & 51         & 0          & 51         & 8             & 1.000             & 3343             & 194             & 0.058        \\
		107     & 50   & 48   & 54         & 49         & 3          & 2             & 0.942             & 3234             & 71              & 0.022        \\
		108     & 48   & 50   & 52         & 47         & 3          & 2             & 0.940             & 3077             & 148             & 0.048        \\
		109     & 45   & 54   & 52         & 1          & 51         & 9             & 0.981             & 4163             & 151             & 0.036        \\
		110     & 50   & 50   & 53         & 49         & 2          & 0             & 0.961             & 3787             & 142             & 0.037        \\
		111     & 64   & 43   & 63         & 59         & 2          & 21            & 0.967             & 4101             & 120             & 0.029        \\
		112     & 51   & 48   & 52         & 49         & 1          & 3             & 0.980             & 3767             & 79              & 0.021        \\
		113     & 57   & 46   & 56         & 54         & 0          & 11            & 1.000             & 3067             & 77              & 0.025        \\
		114     & 44   & 54   & 54         & 1          & 53         & 10            & 0.981             & 3589             & 108             & 0.030        \\ \hline
	\end{tabular}
\end{sidewaystable}

\begin{sidewaystable}\centering
	\caption{Detailed properties and results for House networks}
	\begin{tabular}{lllllllllll}
		\hline
		Session & Year & $n$ & $m$   & density & $m^-$ & $m^+$ & $T(G)$ & $F(G)$ & $L(G)$ & $Y^*$ \\ \hline
		96      & 1979 & 442 & 51081 & 0.524   & 43097 & 7984  & 0.410  & 0.536  & 11845  & 11845 \\
		97      & 1981 & 447 & 49364 & 0.495   & 40304 & 9060  & 0.432  & 0.584  & 10260  & 10259 \\
		98      & 1983 & 444 & 48721 & 0.495   & 36592 & 12129 & 0.511  & 0.569  & 10494  & 10494 \\
		99      & 1985 & 443 & 49764 & 0.508   & 35716 & 14048 & 0.566  & 0.522  & 11885  & 11884 \\
		100     & 1987 & 446 & 50688 & 0.511   & 36780 & 13908 & 0.594  & 0.567  & 10979  & 10979 \\
		101     & 1989 & 449 & 56231 & 0.559   & 39394 & 16837 & 0.606  & 0.565  & 12232  & 12231 \\
		102     & 1991 & 447 & 58067 & 0.583   & 39726 & 18341 & 0.635  & 0.590  & 11914  & 11914 \\
		103     & 1993 & 446 & 59092 & 0.595   & 40290 & 18802 & 0.728  & 0.688  & 9222   & 9222  \\
		104     & 1995 & 445 & 62154 & 0.629   & 44537 & 17617 & 0.777  & 0.797  & 6299   & 6299  \\
		105     & 1997 & 449 & 66701 & 0.663   & 46121 & 20580 & 0.821  & 0.813  & 6238   & 6238  \\
		106     & 1999 & 442 & 63652 & 0.653   & 42753 & 20899 & 0.852  & 0.830  & 5395   & 5395  \\
		107     & 2001 & 447 & 63851 & 0.641   & 43246 & 20605 & 0.859  & 0.848  & 4866   & 4866  \\
		108     & 2003 & 444 & 66277 & 0.674   & 44397 & 21880 & 0.862  & 0.847  & 5057   & 5057  \\
		109     & 2005 & 445 & 66700 & 0.675   & 45420 & 21280 & 0.848  & 0.829  & 5695   & 5695  \\
		110     & 2007 & 452 & 70923 & 0.696   & 47412 & 23511 & 0.853  & 0.842  & 5618   & 5618  \\
		111     & 2009 & 451 & 70160 & 0.691   & 47656 & 22504 & 0.806  & 0.775  & 7877   & 7877  \\
		112     & 2011 & 450 & 77872 & 0.771   & 51084 & 26788 & 0.847  & 0.844  & 6063   & 6063  \\
		113     & 2013 & 447 & 75771 & 0.760   & 48226 & 27545 & 0.890  & 0.880  & 4533   & 4533  \\
		114     & 2015 & 446 & 75180 & 0.758   & 48602 & 26578 & 0.884  & 0.872  & 4801   & 4801  \\ \hline
	\end{tabular}
	\label{tab:housenetworks}
\end{sidewaystable}

\begin{sidewaystable}\centering
	\caption{Democrats (Dems), Republicans (Reps), controlling coalitions (CC), and bill passage in the House}
	\label{tab:housetable}
	\begin{tabular}{lllllllllll}
		\hline
		Session & Dems & Reps & Size of CC & Dems in CC & Reps in CC & Party control & Coalition control & Bills introduced & Signed into law & Passage rate \\ \hline
		96      & 273  & 167  & 244        & 90         & 154        & 106           & 0.631             & 9101             & 477             & 0.052        \\
		97      & 242  & 203  & 262        & 77         & 185        & 39            & 0.706             & 8093             & 317             & 0.039        \\
		98      & 270  & 172  & 258        & 93         & 165        & 98            & 0.640             & 7105             & 392             & 0.055        \\
		99      & 254  & 188  & 236        & 218        & 17         & 66            & 0.928             & 6499             & 381             & 0.059        \\
		100     & 258  & 187  & 251        & 228        & 22         & 71            & 0.912             & 6263             & 458             & 0.073        \\
		101     & 261  & 188  & 241        & 227        & 14         & 73            & 0.942             & 6664             & 387             & 0.058        \\
		102     & 268  & 178  & 295        & 241        & 53         & 90            & 0.820             & 6775             & 411             & 0.061        \\
		103     & 258  & 187  & 248        & 239        & 8          & 71            & 0.968             & 5739             & 301             & 0.052        \\
		104     & 206  & 237  & 253        & 23         & 230        & 31            & 0.909             & 4542             & 253             & 0.056        \\
		105     & 215  & 233  & 235        & 16         & 219        & 18            & 0.932             & 5014             & 262             & 0.052        \\
		106     & 215  & 226  & 227        & 206        & 20         & 11            & 0.912             & 5815             & 410             & 0.071        \\
		107     & 216  & 229  & 224        & 5          & 219        & 13            & 0.978             & 5890             & 312             & 0.053        \\
		108     & 210  & 232  & 227        & 4          & 223        & 22            & 0.982             & 5546             & 356             & 0.064        \\
		109     & 207  & 237  & 229        & 2          & 227        & 30            & 0.991             & 6538             & 332             & 0.051        \\
		110     & 245  & 207  & 244        & 238        & 6          & 38            & 0.975             & 7441             & 314             & 0.042        \\
		111     & 267  & 183  & 254        & 252        & 1          & 84            & 0.996             & 6669             & 265             & 0.040        \\
		112     & 204  & 245  & 249        & 9          & 240        & 41            & 0.964             & 6845             & 205             & 0.030        \\
		113     & 208  & 238  & 236        & 5          & 231        & 30            & 0.979             & 6016             & 219             & 0.036        \\
		114     & 195  & 251  & 242        & 1          & 241        & 56            & 0.996             & 6634             & 218             & 0.033        \\ \hline
	\end{tabular}
\end{sidewaystable}

\FloatBarrier

\subsection*{Movie: Animated versions of coalitions in networks}

Animated versions of the opposing coalitions of the networks is available online at \\ \url{https://saref.github.io/SI/AN2019/Senate_coalitions.mp4}
for the Senate and at \\ \url{https://saref.github.io/SI/AN2019/House_coalitions.mp4}
for the House of Representatives. Looking at the colors of edges we observe that over time, more edges within (between) groups become green (orange) which shows that the networks become more partially balanced and therefore more polarized. If we look at the colors of the nodes (which represent party affiliations), we see that the coalitions become more homogeneous (partisan) over time. These two changes show the two aspects of increase in partisan polarization over time.

\subsection*{Dataset: a-sign-of-the-times.xlsx}
All 38 signed networks used in this study are available as adjacancy matrices stored in an Excel file accessible in a public \textit{Figshare} data repository \cite{Neal2019figshare}. These data are distributed under a CC-BY 4.0 license. This means that one can use these data provided that they provide proper attribution for them by citing the two articles \cite{neal_backbone_2014,neal2018}.

Each tab (sheet) contains the signed network for a chamber of congress indexed using H for House, S for Senate and a number for session (e.g. S98 means the 98th session of the Senate). The first row identifies the representative or senator, with their party affiliation and state; these are square matrices with columns arranged in the same order. In each matrix, a value of $1$ ($-1$) means that the congresspeople associated with that matrix entry have a positive (negative) tie in the network associated with that chamber and session. Likewise, a value of $0$ means that the reciprocal congresspeople have no relationship in that network. Relationships are inferred from bill co-sponsorship data using the Stochastic Degree Sequence Model (SDSM).

The 19 signed networks for the US Senate have slightly over $100$ nodes, each representing a senator, and total edge count between $2073$--$3932$, which results in density values between $0.41$ and $0.78$. The proportions of negative edges are within the range of $0.61$ and $0.82$. The 19 signed networks for the US House of Representatives have slightly over $435$ nodes, each representing a representative, and a total edge count between $48721$ and $77872$, which results in densities in the range of $0.50$--$0.77$. The proportion of negative edges varies between $0.64$ and $0.84$. The networks are slightly larger than the number of legislative seats in their respective chambers because a single seat may be occupied by more than one legislator during a single session, for example due to a death, retirement, or resignation. Accordingly, the nodes in these networks represent individual legislators, not legislative seats.

\subsection*{Dataset: \{House/Senate\}-coalition-membership-results.csv}
The results on globally optimal solutions to the optimization model for computing the frustration index for House and Senate are available in comma-separated values format (two individual csv files) at \\ \url{https://saref.github.io/SI/AN2019/House_coalition_membership_results.csv}
and \\ \url{https://saref.github.io/SI/AN2019/Senate_coalition_membership_results.csv}
.

The first column contains node IDs and the first row contains session numbers. The entry at the intersection of row indexed $r$ and column indexed $c$ represents the optimal value of the $x_i$ variable for a given node (node $r$) in a given session (session $c$).

\end{document}